\documentclass[11pt]{article}
\usepackage{graphicx,latexsym}
\newcommand{\bbox}[1]{\mbox{\boldmath$#1$}}
\begin{document}

\title{Unmixing in random flows}
\author{M. Wilkinson$^1$, B. Mehlig$^2$, S. \" Ostlund$^2$, K. P. Duncan$^1$\\
\\
\\
$^1$Department of Mathematics, The Open University,\\
Walton Hall, Milton Keynes, MK7 6AA, England, \\
$^2$Department of Physics, G\"oteborg University, 41296 G\" oteborg, Sweden.}
\maketitle
\begin{abstract}
We consider particles suspended in a randomly stirred or turbulent
fluid. When effects of the inertia of the particles are
significant, an initially uniform scatter of particles can cluster
together. We analyse this \lq unmixing' effect by calculating the
Lyapunov exponents for dense particles suspended in such a random
three-dimensional flow, concentrating on the limit where the
viscous damping rate is small compared to the inverse correlation
time of the random flow (that is, the regime of large Stokes
number). In this limit Lyapunov exponents are obtained as a power
series in a parameter which is a dimensionless measure of the
inertia. We report results for the first seven orders. The
perturbation series is divergent, but we obtain accurate results
from a Pad\' e-Borel summation. We deduce that particles can
cluster onto a fractal set and show that its dimension is in
satisfactory agreement with previously reported in simulations of
turbulent Navier-Stokes flows. We also investigate the rate of
formation of caustics in the particle flow.
\end{abstract}

\tableofcontents

\section{Introduction}
\label{sec: 0}

\subsection{Overview}
\label{sec: 0.1}

This paper discusses small particles suspended in a randomly
moving fluid. We assume that the fluid flow is mixing, and
concentrate on cases where the fluid flow is incompressible
(although compressibility is considered too, for completeness). At first
sight, it seems as if the particles suspended in an incompressible
mixing flow should become evenly distributed. If the particles
were simply advected by the fluid, this is indeed what would
happen. However, it has been noted that when the effects of finite
inertia of the suspended particles are significant, the particles
can show a tendency to cluster. This remarkable \lq unmixing'
effect was discussed in a theoretical paper by Maxey \cite{Max87},
who proposed that suspended particles (assumed denser than the
fluid) cluster because they are centrifuged away from vortices.
There have been many other theoretical papers on this phenomenon,
and an experimental demonstration was reported by Eaton and
Fessler \cite{Fes94}; the literature is reviewed in section
\ref{sec: 0.2} below. The suspended particles are characterised by
the rate $\gamma $ at which their velocity relative to the fluid
is damped due to viscous drag, and the random velocity field is
characterised by a correlation time $\tau$. The product
$\Omega=\gamma \tau$ is a dimensionless parameter: in much of the
literature ${\rm St}=1/\Omega$ is termed the \lq Stokes number'.
There is a consensus that the clustering effect is observed when
the Stokes number is of order unity.

We investigate a random-flow model with short correlation time,
which is susceptible to mathematical analysis. We show that in
general this model can exhibit pronounced clustering in
circumstances where ${\rm St}\gg 1$, when the centrifuge mechanism
is not effective. When the model is applied to fully-developed
turbulence it predicts that the clustering is strongest when ${\rm
St}=O(1)$ (in agreement with most numerical investigations), but
it indicates that the centrifuge effect is not essential to
understanding the phenomenon.

\begin{figure}
\centerline{\includegraphics[width=12cm]{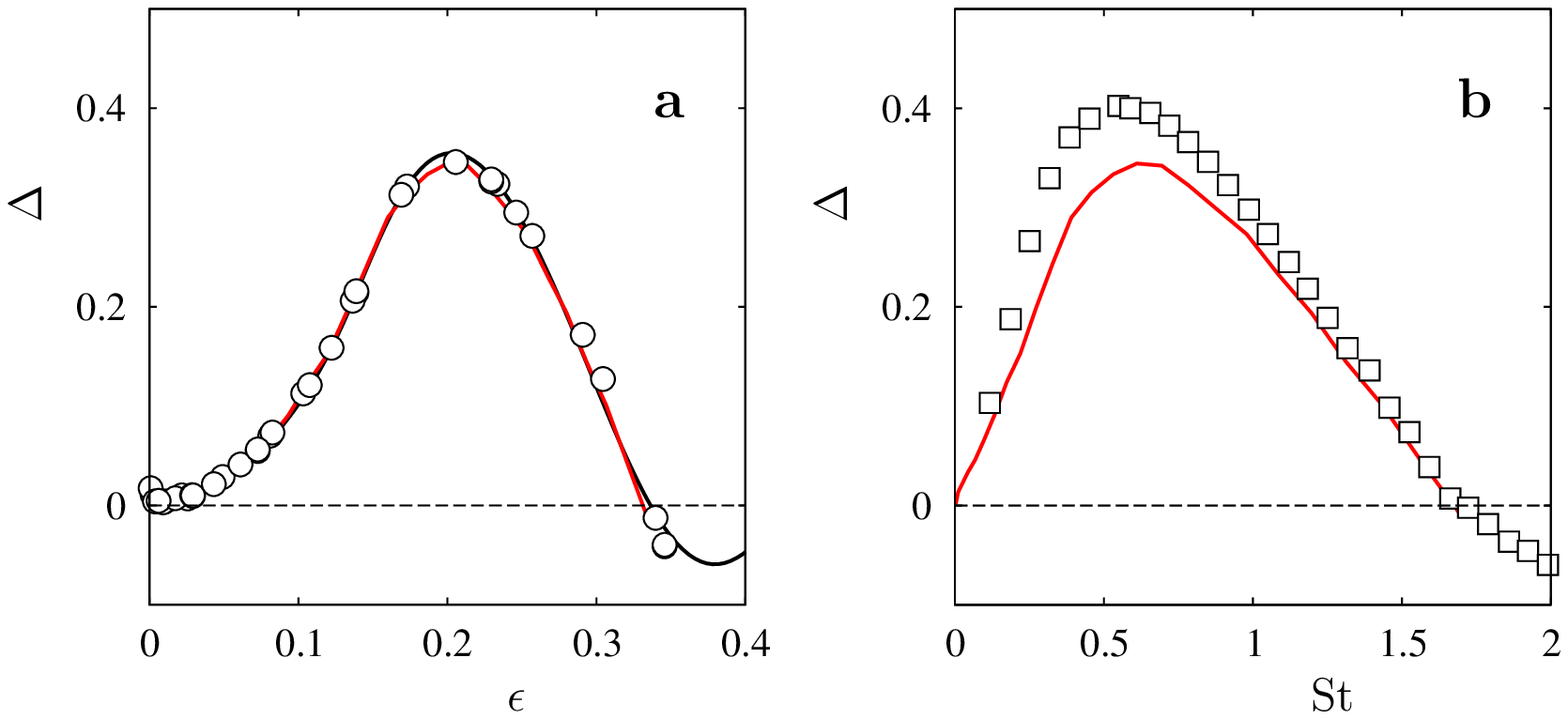}}
\caption{\label{fig:f1} {\bf a} Dimension deficit $\Delta$ as a
function of $\epsilon$, in the limit $\kappa\to 0$; data from
simulation of our short-time correlated random-flow model obtained
as described in section \ref{sec:numcalc} ($\circ$), from a
simulation of the Langevin equation (\ref{eq: 3.14} - \ref{eq:
3.16}), solid red line, and results of a Borel summation of our
perturbation series, using a Pad\'e{} approximant of order
$P_{3/3}$ ($-\!\!\!-\!\!\!-\!\!\!-\!\!\!-\!\!\!-$). The figure
shows that there is pronounced clustering when $\epsilon=O(1)$: in
the limit as $\kappa\to 0$ this happens when ${\rm St}\gg 1$. {\bf
b} Dimension deficit $\Delta$ as a function of Stokes number ${\rm
St}$; results (taken from \cite{Bec06})  of direct  numerical
simulations of particles embedded in Navier-Stokes flows ($\Box$),
and results from Langevin equations of our short-time correlated
random-flow model, (\ref{eq: 3.14} - \ref{eq: 3.16}), solid red
line. The latter are the same as those in figure \ref{fig:f1}{\bf
a}, with ${\rm St}=\epsilon^2/\kappa^2$ and $\kappa=0.256$  was
chosen to give a good fit between the curves, as judged by the
eye. }
\end{figure}

Our principal results are series expansions for the Lyapunov
exponents of the trajectories of the suspended particles in terms
of $\epsilon=\kappa \sqrt{{\rm St}}$, where the dimensionless
parameter $\kappa$ (defined in section \ref{sec: 0.3} below) is
$O(1)$ for fully developed turbulent flow, but may be small for
other types of random flow. Our analysis of the random-flow model
is exact in the limit as $\kappa\to 0$: note that in this limit
${\rm St}\gg 1$ when $\epsilon =O(1)$. We use the Lyapunov
exponents to estimate the \lq dimension deficit' $\Delta$, which
is the difference between the spatial dimension and the (Lyapunov)
fractal dimension $d_{\rm L}$ of the set onto which the particles
cluster: $d_{\rm L}=3-\Delta$ (this will be explained in section
\ref{sec: 0.2}). Figures \ref{fig:f1}{\bf a} and \ref{fig:f1}{\bf
b} illustrate our results. Figure \ref{fig:f1}{\bf a} compares the
value of $\Delta$ as a function of $\epsilon$ obtained by
simulation of our model with results from a Borel summation of our
perturbation series. The results show that there is fractal
clustering when $\epsilon=O(1)$ in the limit as $\kappa\to 0$,
indicating that fractal clustering can occur for large Stokes
numbers. Despite the fact that our perturbation series is
divergent, we obtain satisfactory results from Borel summation.
Figure \ref{fig:f1}{\bf b} compares the results from a direct
numerical simulations of particles in a turbulent Navier-Stokes
flow (these data ($\Box$) are taken from \cite{Bec06}) with
results from our random-flow model: the data used for figure
\ref{fig:f1}{\bf a} are re-plotted with the value of $\kappa $
chosen to give the best fit between the two curves, as judged by
eye. (We cannot determine the scaling theoretically because the
value of $\kappa$ for fully developed turbulence is not known).
The theoretical curve in figure \ref{fig:f1}{\bf b} is obtained
from our random-flow model, which has vanishingly small
correlation time (because for given $\epsilon$, ${\rm St}\to
\infty$ as $\kappa\to 0$). The centrifuge effect cannot therefore
cause clustering in this random-flow model. Our model has a
maximum dimension deficit of approximately $0.35$, as opposed to
$0.40$ for particles in a Navier-Stokes flow, and the form of the
curves is similar. We conclude that the random-flow model provides
a satisfying degree of agreement with full Navier-Stokes dynamics,
despite the fact that the centrifuge mechanism plays no role.

Another mechanism for clustering is the formation of fold caustics
in the flow of the particles. We show that caustics are prevalent
when $\epsilon > 1$. We also consider an explicit expression for
the sum of the Lyapunov exponents when the Stokes number is small.
Taken together with earlier results on the overdamped limit
(referred to in sections \ref{sec: 0.2}, \ref{sec: 7} and
\ref{sec: 8}), our results give a satisfyingly complete
understanding of the local dynamics of suspended particles.

Some of the results were summarised in an earlier letter \cite{Dun05}.

\subsection{Discussion of earlier work}
\label{sec: 0.2}

We first briefly review earlier work on clustering in random
flows, before giving a fuller description of our results.

Maxey \cite{Max87} expressed the trajectory of an inertial
particle in terms of a \lq synthetic' velocity field, which is
obtained  as a perturbation of the velocity field of the fluid. He
showed that this synthetic velocity field has negative divergence
when the vorticity is high or the strain-rate low, and predicted that
particles would have low concentrations in regions of high
vorticity due to this \lq centrifuge effect'. A correlation
between particle density and vorticity has been seen in direct
numerical simulation of particles suspended in a fully-developed
turbulent flow \cite{Wan93,Hog03}.

The mechanism proposed by Maxey led to a prediction concerning the
observability of this \lq preferential concentration' effect. It
is plausible that the \lq centrifuge mechanism' will be less
effective when the particles are overdamped ($\Omega \gg 1$) or
when the velocity fluctuates too rapidly to allow the density of
suspended particles to respond (when $\Omega \ll 1$). This leads
to the hypothesis that the preferential concentration effect
should only be observed when $\Omega$ is close to unity.
Experimental work on particles suspended in fully developed
turbulent air flows \cite{Fes94} appears to support this
hypothesis, as do computer simulations  \cite{Hog03,Gam04,Bec06}.

An alternative approach arises from work of Sommerer and Ott
\cite{Som93}, who discuss patterns formed by particles floating on
a randomly moving fluid. They characterise these patterns in terms
of their fractal dimension, and suggest that the fractal dimension
can be obtained from ratios of Lyapunov exponents $\lambda_i$ of
the particle trajectories, using a formula proposed by Kaplan and
Yorke \cite{Kap79}. The argument of Sommerer and Ott extends
directly to particles suspended in turbulent three-dimensional
flows. The spatial particle flow is characterised by three
Lyapunov exponents, $\lambda_1>\lambda_2>\lambda_3$, and provided
$\lambda_1+\lambda_2>0$ (which is always true for the particles
moving in an incompressible fluid flow), the Kaplan-Yorke estimate
for the fractal dimension is determined by the dimensionless
quantity
\begin{equation}
\label{eq: 0.0} \Delta=-{1\over{\vert \lambda_3
\vert}}(\lambda_1+\lambda_2+\lambda_3)
\end{equation}
which we term the \lq dimension deficit'. When $\Delta >0$, the
Kaplan-Yorke estimate of the dimension, termed the Lyapunov
dimension, is
\begin{equation}
\label{eq: 0.1} d_{\rm L}=3-\Delta
\end{equation}
and $d_{\rm L}=3$ if $\Delta \le 0$. Clustering effects are
significant if the fractal dimension is significantly lower than
the dimension of space. The relations (\ref{eq: 0.0}), (\ref{eq:
0.1}) give a strong motivation to investigate the Lyapunov
exponents of suspended particles. Bec \cite{Bec03,Bec05} has
performed detailed numerical investigations for a specific
ensemble of random flow fields, showing how the Kaplan-Yorke
fractal dimension varies as a function of the Stokes number for
his model flow, reaching a minimum for a value of $\Omega$ which
is of order unity. These numerical calculations of the dimension
deficit are complementary to the ideas proposed by Maxey in
\cite{Max87}, in that they quantify the clustering effect without
explaining its origin.

The motivation for investigating Lyapunov exponents can also be
explained without referring to the fractal dimension. For
three-dimensional flows, the sum of the three largest Lyapunov
exponents of the particle trajectories may be negative, implying
that volume elements almost always contract. However, for
incompressible flows the first Lyapunov exponent is always
positive, implying that nearby particles almost surely separate.
If $(\lambda_1+\lambda_2+\lambda_3)/\lambda_1$ is negative, there
is a tendency for particles to cluster, but if the magnitude of
this quantity is small compared to unity the clustering effect
will be negligible, because in that case clusters are stretched
and folded more rapidly than their density accumulates.

Most of the theoretical work on clustering in turbulent flows has
emphasised instantaneous correlations between vortices and
particle-density fluctuations. There is an alternative viewpoint
on the origin of density fluctuations which is also theoretically
tenable, and which is the basis for our own analysis. It can be
argued that the density fluctuations are generated by a
multiplicative random process: volume elements in the particle
flow are randomly compressed or expanded, and the ratio of the
final density to the initial density after many multiples of the
correlation time $\tau$ can be modeled as a product of a large
number of random factors. According to his picture, the density
fluctuations will be a record of the history of the flow, and may
bear no relation to the instantaneous disposition of vortices when
the particle density is measured. The particle density is expected
to have a log-normal probability distribution, and the mean value
of the logarithm is related to the Lyapunov exponents of the flow.
This is another motivation for calculating Lyapunov exponents.

The idea of considering clustering due to random flows has been
used in earlier work which adapted results relating to purely
advective flows. Most of this literature treats the limiting case
where the correlation time of the velocity field is very short, a
case which is known as the Kraichnan model \cite{Kra74}. The
Lyapunov exponents of such advective flows have been calculated in
different ways by several authors: these calculations include
results for compressible and solenoidal (incompressible) flows:
the earliest calculation appears to be by LeJan \cite{LeJ85}, who
treated a spatially correlated Brownian motion. Later work
\cite{Ber00,Fal01} subsequently showed that the same results apply
to a flow with a smooth time dependence, but a very short
correlation time. These calculations on purely advective flows
cannot explain clustering of particles suspended in incompressible
flows, because the density of advected particles remains constant
for incompressible flow. Elperin \cite{Elp96} proposed to analyse
the motion of inertial particles by applying
results derived for
advective flows to the synthetic velocity field derived by Maxey
\cite{Max87}, which has a compressible component. The same
approach was subsequently adopted in \cite{Bal01}. The
approximations employed in these papers make use of results from
the Kraichnan model of passive scalars \cite{Kra74}, valid in the
limit where the correlation time $\tau$ approaches zero (St
$\rightarrow\infty$). Yet Maxey's synthetic velocity field is
obtained in the overdamped limit where, by contrast,  the Stokes
number is small, St $\rightarrow 0$.

In summary, the clustering effect can be characterised by
calculating the Lyapunov exponents of the particle trajectories,
but there is a limited theoretical understanding of these, based
on Maxey's correction to advective flow (which is derived in the
limit where $\Omega \ll 1$). There is a consensus that significant
clustering only occurs when $\Omega \approx 1$ but there is scope
for revising this expectation.

Our own work \cite{Wil03,Meh04,Wil05,Meh05,Dun05} uses a model for
a random flow with a very short correlation time, but the effects
of particle inertia are properly accounted for. We remark that a
similar approach was proposed by Piterbarg \cite{Pit01}, who
studied the largest Lyapunov exponent in a two-dimensional flow.

\subsection{Plan of paper and summary of results}
\label{sec: 0.3}

In section \ref{sec: 1}, we discuss the dimensionless parameters
of the model, and in particular we note the significance of an
additional dimensionless parameter, the \lq Kubo number' (this
term is used in the plasma-physics literature, \cite{Bri74}). In
terms of a typical fluid velocity $u$ and correlation length
$\xi$, the Kubo number is $\kappa=u\tau/\xi$. We argue that
$\kappa$ cannot be large, that $\kappa=O(1)$ for fully-developed
turbulent flows, and that $\kappa$ may be small in other systems,
such as randomly stirred fluids.

Most of our paper is concerned with the case where the Stokes
number ${\rm St}=1/\Omega$ is large. This is the underdamped limit
where inertial effects are most likely to be significant and where
very little work has been done previously. In section \ref{sec: 2}
we show how the three Lyapunov exponents for inertial particles
can be obtained from expectation values of a system of nine
coupled stochastic differential equations.

In section \ref{sec: 3} we discuss how these general equations can
be represented by a system of Langevin equations in the limit
where the Stokes number is large and the Kubo number is small. A
dimensionless parameter $\epsilon\propto\kappa\,\Omega^{-1/2}$
plays a natural role in these Langevin equations: inertial effects
are significant when $\epsilon$ is large.

Section \ref{sec: 10} shows how the corresponding Fokker-Planck
equation can be mapped to a perturbation of a nine-dimensional
isotropic quantum harmonic oscillator. Using the algebra of
harmonic-oscillator raising and lowering operators, we develop
perturbation expansions for the Lyapunov exponents to large orders
in $\epsilon$, with exact expressions for the coefficients. These
perturbation expansions are presented in section \ref{sec: 6} and
are the principal results in this paper.

In the case where inertia is important it is possible for the
density to diverge due to the formation of caustics, which are
surfaces on which the Jacobian determinant of the particle flow
field vanishes. In section \ref{sec: 6} we also quantify the rate
of formation of these caustic surfaces. Caustics can have a very
significant effect on the aggregation of suspended particles,
because they can produce a divergent density of particles in a
finite time \cite{Wil05} and because they greatly increase the
relative velocity of suspended particles \cite{Wil06}.

The perturbation series are divergent, and in section \ref{sec: 6}
we also discuss their summation by Borel-Pad\'e methods. We find
excellent agreement with numerical simulations in some cases, but
there are revealing discrepancies in other cases: some of our
Borel-Pad\'e summations differ from Monte-Carlo evaluations by
quantities which have a non-analytic dependence on the
perturbation parameter $\epsilon$, of the form
$\exp(-\Phi^\ast/\epsilon^2)$ (for some constant $\Phi^\ast$). In
section \ref{sec: WKB} we consider a WKB approach to solving the
Fokker-Planck equation, and indicate how such non-analytic
contributions arise.

Section \ref{sec: 7} makes connections between our results and
earlier work on Lyapunov exponents of advected particles: we show
that the leading-order term in our perturbation series agrees with
the Lyapunov exponent for advective flow, and we show that in the
limit as $\kappa\to 0$ the same expression for the Lyapunov
exponent holds, regardless of whether $\Omega$ is large or small.
In the case where the flow is incompressible the sum of the
Lyapunov exponents for advected particles is zero. In section
\ref{sec: 8} we calculate the leading contribution due to the sum
of the Lyapunov exponents for particles in an incompressible flow
when $\Omega\gg 1$ and $\kappa\ll 1$: the results confirm that
$d_{\rm L}$ is very close to $3$ when $\Omega \gg 1$.

Our results have important   consequences for the theory of
particle clustering. The existing consensus favours the view that
clustering only occurs when $\Omega\sim 1$, and is the result of
the \lq centrifuge effect'. Our results present a different
picture. Clustering onto a fractal set occurs when $\Delta $,
defined by (\ref{eq: 0.0}), is positive, and becomes significant
when this number is of order unity. Both simulations and the
Borel-Pad\' e summations in section \ref{sec: 6} indicate that
when $\Omega\ll 1$ and $\kappa\ll 1$, $\Delta $ is positive when
$\epsilon $ is less than a critical value $\epsilon_{\rm c}$, and
achieves its maximum value of $\Delta_{\rm max}\approx 0.35$ when
$\epsilon$ equals $\epsilon_{\rm max}\approx 0.64 \epsilon_{\rm
c}$. Thus we establish that clustering can be significant when
$\Omega\ll 1$, provided that $\kappa\ll 1$. Our results on the
overdamped limit obtained in section \ref{sec: 8} indicate that
although $\Delta>0$ when $\Omega\gg 1$, it is very small, implying
that clustering effects are hard to observe when $\Omega\gg 1$.
For fully developed turbulence we have $\kappa\sim 1$, which is on
the border of the region of validity of our theory, but a plot of
$\Delta $ versus ${\rm St}$ for our model (figure \ref{fig:f1}{\bf
b}) shows satisfying agreement with numerical simulations of
turbulent flows (after scaling ${\rm St}$ to account for
uncertainties in the definitions of correlations times). This
indicates that the centrifuge mechanism makes a marginal
contribution to the clustering process. Systems such as randomly
stirred fluids, or particles falling under gravity through fully
developed turbulence, can exhibit flows where $\kappa\ll 1$, and
may exhibit clustering for large values of ${\rm St}$.

\section{Formulation of the problem}
\label{sec: 1}

\subsection{Equations of motion}
\label{sec: 1.1}

We assume that the particles suspended in the fluid flow satisfy
the equations of motion
\begin{equation}
\label{eq: 1.1}
\dot {\mbox{\boldmath$r$}}={1\over {m}}\mbox{\boldmath$p$}
\ ,\ \
\dot {\mbox{\boldmath$p$}}=m\gamma \bigl[\mbox{\boldmath$u$}
(\mbox{\boldmath$r$},t)
-\dot {\mbox{\boldmath$r$}}\bigr]
\end{equation}
where $\bbox{r} = (r_1,r_2,r_3)$ is the position of a particle,
$\mbox{\boldmath$p$}$ is the particle momentum, $m$ is its mass
and $\bbox{u}(\bbox{r},t)$ denotes the velocity field. We neglect
effects due to the inertia of the displaced fluid: this is
justified when $\rho_{\rm p}/\rho_{\rm f}\gg 1$, where $\rho_{\rm
f}$, $\rho_{\rm p}$ are the densities of the fluid and particles
respectively. Equations (\ref{eq: 1.1}) are appropriate for
spherical particles when the Reynolds number of the flow referred
to the particle diameter is small, and when the particle radius
$a$ satisfies $a\ll \xi$. Further conditions are required for the
validity of this formula: these can always be satisfied if the
radius of the particle and the molecular mean free path of the
fluid are sufficiently small \cite{Max83}.

Stokes's formula gives the relaxation rate
\begin{equation}
\label{eq: 1.2} \gamma={6\pi a\rho_{\rm
f}\nu\over{m}}=\frac{9\rho_{\rm f}\nu}{2\rho_{\rm p}a^2}
\end{equation}
where $\nu$ is the kinematic viscosity of the fluid. The neglect
of the mass of the displaced fluid is an excellent approximation
for aerosol systems, and satisfactory for many examples of solid
particles in water.

We also assume that the effect of Brownian diffusion of the
particles is negligible: the ratio of the particle diffusion
constant to the molecular diffusion constant of the fluid is
proportional to the ratio of the molecular mean free path to the
particle diameter.

\subsection{Dimensionless parameters}
\label{sec: 1.2}

We characterise the random velocity field by its statistics,
denoting the expectation value of a quantity $X$ by $\langle
X\rangle$. We assume that the mean velocity is zero: $\langle
\mbox{\boldmath$u$}(\mbox{\boldmath$r$},t)\rangle={\bf 0}$. The
velocity field can be characterised by its correlation function,
which has a correlation length $\xi$ and a correlation time
$\tau$. The fluctuations of the fluid velocity have a
characteristic scale $u$: in a single-scale flow we would define
$u^2=\langle \mbox{\boldmath$u$}^2\rangle$, but in fully developed
turbulence it is more natural to define $u$ as the velocity scale
associated with fluctuations on the dissipative scale, $u=({\cal
E} \nu)^{1/4}$, where ${\cal E}$ is the rate of dissipation per
unit mass \cite{Fri97}.

The equations of motion (\ref{eq: 1.1}) are characterised by four
dimensional parameters. The random velocity field is described by
three scales: $u$, $\xi$ and $\tau$. In addition, the interaction
of the fluid with the particles is described by the damping rate
$\gamma$. (The mass $m$ can be eliminated from the two components
of (\ref{eq: 1.1}), but may appear in expressions which contain
forces). From the four quantities $\xi,\tau,u,\gamma$ we can form
two independent dimensionless groups, the Kubo number, $\kappa$
and the Stokes number, $1/\Omega$. The degree to which the
velocity field is compressible is described by a further
dimensionless variable, $\Gamma$, which will be defined in section
\ref{sec: 3}. The average number of suspended particles per unit
volume, $N_0$, is associated with a further independent
dimensionless parameter, $\Upsilon=N_0\xi^3$. The set of
dimensionless parameters of the system is therefore
\begin{equation}
\label{eq: 1.3} \kappa={u\tau\over{\xi}} \ ,\ \ \ \Omega=\gamma
\tau \ ,\ \ \ \Gamma \ ,\ \ \ \Upsilon=N_0\xi^3\,.
\end{equation}
We consider flow fields ranging from incompressible flow (which
corresponds to $\Gamma=2$, see equation (\ref{eq:gamma}) in
section \ref{sec: 3}) to pure potential flow ($\Gamma ={1\over
3}$). We concentrate on the underdamped limit ($\Omega \ll 1$,
large Stokes number), but also give results for the overdamped
situation.

We argue on physical grounds that $\kappa $ cannot be large, and
that $\kappa \sim 1$ for fully developed turbulence. For real
flows satisfying the Navier-Stokes equations, the velocity field
is self-convected, so that temporal variations of the velocity
field at any point are partly due to the \lq sweeping' action of
the flow. If $u$ is the characteristic velocity and $\xi$, $\tau$
are the spatial and temporal correlation scales, then the
transport of the velocity field by its own action will cause it to
fluctuate on a time scale $\xi/u$, which cannot be less than the
actual correlation time of the field. Thus $\kappa=u\tau/\xi$
cannot be large. Small Kubo numbers are realised in randomly
stirred fluids where the Reynolds number is small enough that the
flow does not spontaneously generate turbulence. For fully
developed turbulence, the Kolmogorov theory \cite {Fri97}
indicates that $u$, $\xi$ and $\tau$ are all functions of ${\cal
E}$ and $\nu$, and dimensional considerations imply that
$\kappa\sim 1$. Our analytical results are all derived in the
limit where $\kappa \ll 1$, so that fully-developed turbulence is
on the borderline of applicability of our theory.

A practically important measure of the degree of clustering is
given by the ratio of the largest observed particle density
$N_{\rm max}$ to the mean particle density. The parameter
$\Upsilon $ will have a pronounced effect on the distribution
pattern of the particles. When $\Upsilon$ is sufficiently small,
the particles will appear as a random scatter if the largest
Lyapunov exponent (defined below) is positive. The set of particle
positions is a point set which randomly samples a fractal measure,
but the fractal is only visible when $\Upsilon $ is sufficiently
large. For large $\Upsilon$, the density-enhancement factor
$N_{\rm max}/N_0$ may be very large, even when the parameter
$\Delta $ (defined by (\ref{eq: 0.0})) is small. A quantitative
discussion of these issues would be quite lengthy. Accordingly, in
this paper we will consider only locally defined properties of the
particle trajectories, namely the Lyapunov exponents and the rate
of caustic formation, both of which are defined below.

We remark that the dimensionless parameters $\kappa$ and
$\Upsilon$ have apparently not been considered in earlier papers
on clustering of particles in random flow fields. Uncertainties
about the intended values of these parameters makes some of the
literature quite difficult to understand.

\subsection{Definitions of rate constants}
\label{sec: 1.3}

The Lyapunov exponents $\lambda_i$, $i=1,2,3$ are rate constants
which are defined in terms of the time dependence of small
separations of trajectories from a reference trajectory
$\mbox{\boldmath$r$}(t)$. We consider three trajectories which
have infinitesimal displacements from the reference trajectory,
$\delta\mbox{\boldmath$r$}_i(t)$, $i=1,2,3$. We then consider the
length $\delta r=\vert \delta\mbox{\boldmath$r$}_1\vert$ of a
small separation between two trajectories, the area $\delta {\cal
A}=\vert\mbox{\boldmath$r$}_1\wedge \mbox{\boldmath$r$}_2\vert$ of
a parallelogram spanned by two separation vectors and the volume
$\delta {\cal
V}=\vert\mbox{\boldmath$r$}_3\cdot\mbox{\boldmath$r$}_2\wedge\mbox{\boldmath$r$}_3\vert$
of a parallelepiped spanned by a triad of separations. The
Lyapunov exponents are defined by writing
\begin{eqnarray}
\label{eq: 2.4}
\lambda_1&=&\lim_{t\to \infty}{1\over t}\log_{\rm e}(\delta r)
\nonumber \\
\lambda_1+\lambda_2&=&\lim_{t \to \infty}{1\over t}\log_{\rm e}(\delta{\cal A})
\nonumber \\
\lambda_1+\lambda_2+\lambda_3&=&\lim_{t \to \infty}{1\over t}
\log_{\rm e}(\delta{\cal V})
\ .
\end{eqnarray}
If $\lambda_1<0$, pairs of particles coalesce with probability
unity. If $\lambda_1>0$ and $\lambda_1+\lambda_2<0$, particles
cluster onto randomly moving lines, which stretch and fold. If
$\lambda_1+\lambda_2>0$ but $\lambda_1+\lambda_2+\lambda_3<0$, the
particles cluster onto randomly stretching and folding surfaces.

\subsection{Caustics}
\label{sec: 1.4}

\begin{figure}
\centerline{\includegraphics[width=8cm]{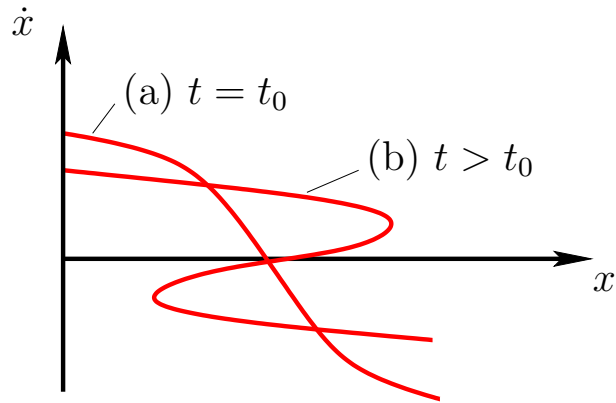}}
\caption{\label{fig:c} Schematic illustration of a caustic in one
spatial dimension: the velocity $\dot x$ of the particles as a
function of position $x$ is initially single-valued (curve (a))
but particles with a large velocity overtake slower-moving
particles, so that at a later time the particle velocity is
multi-valued (curve (b)). The region where the velocity is a
multi-valued function is bounded by two fold caustics.  }
\end{figure}

Another locally defined statistic is the rate of caustic
formation. There is no constraint which prevents two of the three
vectors defining the separation between nearby particles becoming
collinear, so that the volume element $\delta {\cal V}$ becomes
zero for an instant in time. These events correspond to \lq
caustics', where faster moving particles overtake slower ones
\cite{Fal02,Wil05,Wil06} (see figure \ref{fig:c} for an
illustration in one spatial dimension). Caustics influence the
spatial particle distribution and the relative velocities of
nearby particles.

There is an increased density of particles on the fold caustics
(which are a pair of points in the one-dimensional example of
figure \ref{fig:c}, but which form surfaces in three dimensions),
and the particle density on the caustics diverges in the limit as
$\Upsilon \to \infty$. This effect is discussed in
\cite{Fal02,Wil05}: it is analogous to the divergence of light
intensity on optical caustics \cite{Ber81}.

The other effect of caustics is that the particle velocity field
becomes multi-valued in the region between the caustics (in figure
\ref{fig:c} the velocity field is triple-valued between the
caustics). Because particles at the same position are moving with
differing velocities, the caustics enhance the rate of collision
of suspended particles \cite{Fal02,Wil05} (this has no analogue in
optical caustics). Caustics therefore facilitate the aggregation
of suspended particles.

We define $J$, the rate of caustic formation, as the rate at which
events where $\delta {\cal V}=0$ occur for a given triplet of
nearby trajectories. We define a dimensionless rate $J'$ by $J' =
J/\gamma$.

\section{Equations determining the Lyapunov exponents}
\label{sec: 2}

\subsection{Stochastic differential equations for the Lyapunov
exponents}
Linearising the equations of motion (\ref{eq: 1.1}) gives
\begin{eqnarray}
\label{eq: 2.1}
\delta \dot {\mbox{\boldmath$p$}}&=&
-\gamma \delta \mbox{\boldmath$p$}+{\bf F}(t)
\delta \mbox{\boldmath$r$}
\nonumber \\
\delta \dot {\mbox{\boldmath$r$}}&=&{1\over m}\delta \mbox{\boldmath$p$}
\ .
\end{eqnarray}
where ${\bf F}(t)$ is a matrix with elements proportional to the
rate-of-strain matrix:
\begin{equation}
\label{eq: 2.2}
F_{\mu\nu}(t)=\gamma m\, {\partial u_\mu\over{\partial r_\nu}}
(\mbox{\boldmath$r$}(t),t)
\end{equation}
(from this point we will use Greek subscripts to label components
in three dimensional space, reserving Roman indices for components
in a nine-dimensional space which appears later). To determine the
Lyapunov exponents we consider three trajectories displaced
relative to a reference trajectory by $(\delta
\mbox{\boldmath$r$}_\mu,\delta \mbox{\boldmath$p$}_\mu)$, with
$\mu=1,2,3$. We choose to parametrise the spatial displacements as
follows
\begin{eqnarray}
\label{eq: 2.3}
\delta \mbox{\boldmath$r$}_1&=&X_1{\bf n}_1
\nonumber \\
\delta \mbox{\boldmath$r$}_2&=&X_2({\bf n}_1+\delta \theta{\bf n}_2)
\nonumber \\
\delta \mbox{\boldmath$r$}_3&=&X_3[{\bf n_1}+\delta \theta \delta \phi
(Z{\bf n}_2+{\bf n}_3)]
\end{eqnarray}
where the ${\bf n}_\mu(t)$ form a triplet of orthogonal unit unit
vectors. Nine variables are required to parametrise the spatial
displacements $\delta \mbox{\boldmath$r$}_\mu$. Two parameters
specify the direction of ${\bf n}_1$, and a further angular
parameter specifies the direction of ${\bf n}_2$ relative to ${\bf
n}_1$. There is only a binary choice in the direction of ${\bf
n}_3$, and we resolve this by requiring continuity. This means
that a further six parameters are required, and (\ref{eq: 2.3})
does indeed contain a further six parameters, namely
$(X_1,X_2,X_3,Z,\delta \theta,\delta \phi)$.

For the choice of parametrisation above, the small elements
required in (\ref{eq: 2.4}) are $\delta r=X_1$, $\delta {\cal
A}=X_1X_2\delta \theta$ and $\delta {\cal V}=X_1X_2X_3\delta
\theta^2 \delta \phi$. We can then extract the Lyapunov exponents
from expectation values of logarithmic derivatives of the
variables used in our parametrisation:
\begin{eqnarray}
\label{eq: 2.5} \lambda_1&=&\biggl\langle {\dot
X_1\over{X_1}}\biggr\rangle\nonumber\\
\lambda_2&=&\biggl\langle {\dot X_2\over{X_2}}\biggr\rangle
+\biggl\langle{\delta \dot \theta\over{\delta \theta}}\biggr\rangle
\nonumber \\
\lambda_3&=&\biggl\langle {\dot X_3\over{X_3}}\biggr\rangle
+\biggl\langle {\delta \dot \phi\over{\delta \phi}}\biggr\rangle
+\biggl\langle {\delta \dot \theta\over{\delta \theta}}\biggr\rangle
\ .
\end{eqnarray}
The angles $\delta \theta$ and $\delta \phi$
both decrease with probability unity, and
eventually we retain only leading orders in these
variables. Initially, however, we retain all
terms.

The equations of motion for the displacements
$\delta \mbox{\boldmath$r$}_\mu$
also contain corresponding increments of momentum,
$\delta \mbox{\boldmath$p$}_\mu$. These will be parametrised in terms
of the nine elements of a $3\times 3$ matrix $\bf R$ by writing
\begin{equation}
\label{eq: 2.6}
\delta \mbox{\boldmath$p$}_\mu= {\bf R} \,\delta \bbox{r}_\mu
\ .
\end{equation}
Substituting this relation into (\ref{eq: 2.1}) gives the
equation of motion for $\bf R$:
\begin{equation}
\label{eq: 2.7}
{{\rm d}{\bf R}\over{{\rm d}t}}=-\gamma {\bf R}-{1\over m}{\bf R}^2+
{\bf F}(t)
\ .
\end{equation}
Now we determine the equations of motion for
the parameters determining $\delta \mbox{\boldmath$r$}$.
Differentiating each line of (\ref{eq: 2.3}) with respect to
time, and then using the second equation of (\ref{eq: 2.1}), we
obtain
\begin{eqnarray}
\label{eq: 2.8} &&
\dot X_1{\bf n}_1+X_1\dot {\bf n}_1\nonumber\\
&&\hspace*{1cm}={1\over m}X_1{\bf R}(t){\bf n}_1
\nonumber \\[2mm]
&&\dot X_2({\bf n}_1+\delta \theta{\bf n}_2)
+X_2(\dot {\bf n}_1+\delta \theta \dot {\bf n}_2)+X_2\delta \dot \theta
{\bf n}_2
\nonumber \\
&&\hspace*{1cm}={1\over m}X_2{\bf R}(t)({\bf n}_1+\delta \theta
{\bf n}_2)
\nonumber \\[2mm]
&&\dot X_3[{\bf n}_1+\delta \theta
\delta \phi(Z{\bf n}_2+{\bf n}_3)]+X_3[\dot {\bf n}_1+\delta \theta
\delta \phi(Z\dot {\bf n}_2+\dot {\bf n}_3)]
\nonumber \\
&&\hspace*{1cm}+X_3(\delta \dot \theta \delta \phi
+\delta \theta \delta \dot \phi)
(Z{\bf n_2}+{\bf n}_3)+X_3\delta \theta \delta \phi \dot Z{\bf n}_2
\nonumber \\
&&\hspace*{1cm}={1\over m}X_3{\bf R}(t)[{\bf n}_1+\delta \theta
\delta \phi (Z{\bf n}_2+{\bf n}_3)] \ .
\end{eqnarray}
We now take the scalar product of each of these three
equations with each of the unit vectors ${\bf n}_\mu$ in turn.
It is convenient to use the notations
\begin{equation}
\label{eq: 2.9} R'_{\mu\nu}(t)={\bf n}_\mu (t)\cdot{\bf
R}(t)\,{\bf n}_\nu(t) \ ,\ \ \ F'_{\mu\nu}(t)={\bf n}_\mu
(t)\cdot{\bf F}(t)\,{\bf n}_\nu(t)
\end{equation}
for the elements of the tensors ${\bf F}$ and ${\bf R}$ transformed
to the rotated basis. Taking the scalar product of the
first equation of (\ref{eq: 2.8}) with each of the
unit vectors leads to the equations
\begin{equation}
\label{eq: 2.10}
{\dot X_1\over{X_1}}={1\over m}R'_{11}(t)
\end{equation}
\begin{equation}
\label{eq: 2.11}
\dot {\bf n}_1\cdot {\bf n}_2={1\over m}R'_{21}(t)
\end{equation}
\begin{equation}
\label{eq: 2.12}
\dot {\bf n}_1\cdot {\bf n}_3={1\over m}R'_{31}(t)
\ .
\end{equation}
Now taking the scalar product of the second equation
of (\ref{eq: 2.8}) with each unit vector in turn, and using
(\ref{eq: 2.10}) to (\ref{eq: 2.12}) to simplify gives, respectively
\begin{equation}
\label{eq: 2.13}
{\dot X_2\over{X_2}}={\dot X_1\over{X_1}}+{1\over m}\delta \theta
[R'_{12}(t)+R'_{21}(t)]
\end{equation}
\begin{equation}
\label{eq: 2.14} {\delta \dot \theta\over{\delta \theta}}={1\over
m}[R'_{22}(t)-R'_{11}(t)] -{\delta
\theta\over{m}}[R'_{12}(t)+R'_{21}(t)]
\end{equation}
\begin{equation}
\label{eq: 2.15}
\dot {\bf n}_2 \cdot {\bf n}_3={1\over m}R'_{32}(t)\,.
\end{equation}
Using the final equation of (\ref{eq: 2.8}), and
making use of (\ref{eq: 2.10}) to (\ref{eq: 2.15}) to
simplify, we find
\begin{equation}
\label{eq: 2.16}
{\dot X_3\over{X_3}}={\dot X_1\over{X_1}}+{1\over m}\delta \theta
\delta \phi
\biggl(Z[R'_{12}(t)+R'_{21}(t)]+R'_{13}(t)+R'_{31}(t)\biggr)
\end{equation}
\begin{equation}
\label{eq: 2.17}
{\delta \dot \phi\over{\delta \phi}}={1\over m}[R'_{33}(t)-R'_{22}(t)]
+{\delta \theta\over {m}}[R'_{12}(t)+R'_{21}(t)]
-{\delta \theta \delta \phi\over{m}}
\biggl(Z[R'_{12}(t)+R'_{21}(t)]+R'_{13}(t)+R'_{31}(t)\biggr)
\end{equation}
\begin{equation}
\label{eq: 2.18}
\dot Z={1\over m}[R'_{23}(t)+R'_{32}(t)]+{1\over m}Z[R'_{22}(t)-R'_{33}(t)]\,.
\end{equation}
Equations (\ref{eq: 2.10}) to (\ref{eq: 2.18}) are the exact
equations of motion for the nine variables parametrising
$\delta \mbox{\boldmath$r$}$. Retaining only the leading-order terms in $\delta \theta$ and $\delta \phi$, we
have
\begin{equation}
\label{eq: 2.19}
{\dot X_1\over X_1}={1\over m}R'_{11}
\ ,\ \ \
{\delta \dot \theta\over{\delta \theta}}={1\over m}(R'_{22}-R'_{11})
\ ,\ \ \
{\delta \dot \phi \over{\delta \phi}}={1\over m}(R'_{33}-R'_{22})\,.
\end{equation}
Using equation (\ref{eq: 2.5}), we find the following expressions
for the Lyapunov exponents:
\begin{equation}
\label{eq: 2.20} \lambda_1={1\over m}\langle R'_{11}\rangle \ ,\ \
\ \lambda_2={1\over m}\langle R'_{22}\rangle \ ,\ \ \
\lambda_3={1\over m}\langle R'_{33}\rangle \ .
\end{equation}
The Lyapunov exponents are thus obtained directly from the
elements of $\langle{\bf R}'\rangle$ which satisfies an equation
similar to (\ref{eq: 2.7}): to derive it, write ${\bf n}_\mu =
{\bf O}\, {\bf e}_\mu$ where ${\bf e}_\mu$ are the unit vectors of
a fixed Cartesian coordinate system, ${\bf n}_\mu$ were introduced
in (\ref{eq: 2.3}), and ${\bf O}$ is an orthogonal matrix.
Transforming (\ref{eq: 2.7})
\begin{equation}
\label{eq: 2.23}
{\bf O}^+ \dot{\bf R} {\bf O} = -\gamma {\bf R}' -\frac{1}{m} {{\bf R}'}^2
+{\bf F}'\,.
\end{equation}
Making use of
\begin{equation}
\label{eq: 2.24}
{\bf O}^+ \dot{\bf R} {\bf O} = \dot {{\bf R}'} -[{\bf R}',
{\bf O}^+ \dot{\bf O}]
\end{equation}
we  obtain
\begin{equation}
\label{eq: 2.25}
\dot {{\bf R}'} = -\gamma {\bf R}' -\frac{1}{m} {{\bf R}'}^2
+[{\bf R}',{\bf O}^+ \dot{\bf O}]
+{\bf F}'\,.
\end{equation}
The matrix elements of ${\bf O}^+ \dot{\bf O}$ are given by
(\ref{eq: 2.11}), (\ref{eq: 2.12}), and (\ref{eq: 2.15})
\begin{equation}
\label{eq: 2.26}
{\bf O}^+ \dot{\bf O}=\frac{1}{m}
\left( \begin{array}{lcr}
0       &-R_{21}^\prime&-R_{31}^\prime\\
R_{21}' &      0 &-R_{32}'\\
R_{31}' & R_{32}'& 0
\end{array}\right)\,.
\end{equation}
To summarise: the Lyapunov exponents are determined by (\ref{eq: 2.20}),
using expectation values of variables occurring in the
system of stochastic differential equations defined by (\ref{eq: 2.25}) and (\ref{eq: 2.26}).

\subsection{Numerical calculation of Lyapunov exponents}
\label{sec:numcalc}

\begin{figure}
\centerline{\includegraphics[width=12cm]{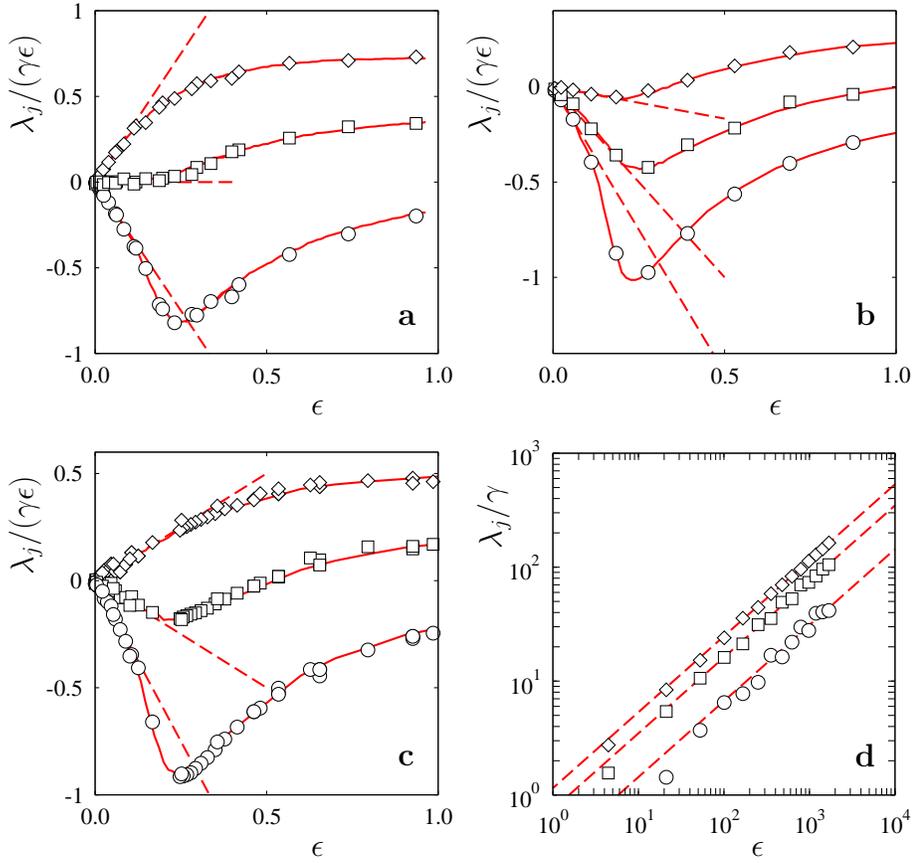}}
\caption{\label{fig:f2} Lyapunov exponents $\lambda_j$ as a
function of $\epsilon$ for $j = 1,2,3$ and three different values
of $\Gamma$. Panels {\bf a} - {\bf c} show numerical simulations
using a discretised form of the equations of motion (\ref{eq:
1.1}), replacing the force field by a random variable with
appropriate statistics as described in section \ref{sec:numcalc}:
$\diamond$ ($j=1$), $\Box$ ($j=2$), and $\circ$ ($j=3$). Also
shown are results of Monte-Carlo simulation of the Langevin
equation (\ref{eq: 3.14}), solid lines, as well as the
lowest-order estimates (\ref{eq:underdamped}), dashed lines. Panel
{\bf a}: $\Gamma=2$ (incompressible flow). Panel {\bf b}:
$\Gamma=1/3$ (pure potential flow). Panel {\bf c}: $\Gamma=1$.
Panel {\bf d} shows large-$\epsilon$ behaviour of the numerical
data from {\bf a}, also shown are fits to the large-$\epsilon$
behaviour (\ref{eq: 3.17}), dashed lines.}
\end{figure}

Our numerical results for the Lyapunov exponents, represented as
symbols in figures \ref{fig:f1}, \ref{fig:f2}, \ref{fig:f4}, and
\ref{fig:f5}, were obtained in the limit of $\tau \rightarrow 0$
by discretising time in equation (\ref{eq: 2.1}). We write $t =
n\delta t$  with $n = 0,1,2,\ldots$, and with a time step $\delta
t$ which satisfies $1/\gamma\gg \delta t\gg\tau$. Writing $\delta
\bbox{r}_n = \delta \bbox{r}(n \delta t)$ and $\delta \bbox{p}_n =
\delta \bbox{p}(n \delta t)$ the Euler discretisation of (\ref{eq:
2.1}) is
\begin{equation}
\left(\begin{array}{c}
\delta \bbox{r}^{(n+1)} \\
\delta \bbox{p}^{(n+1)}
\end{array}\right)
=
\left(\begin{array}{ll}
{\bf I} & {\bf I}\,(\delta t/m)\\
{\bf F}^{(n)} \delta t & {\bf I}\, (1-\gamma \delta t)
\end{array}\right)
\left(\begin{array}{c}
\delta \bbox{r}^{(n)} \\
\delta \bbox{p}^{(n)}
\end{array}\right)
\equiv {\bf M}^{(n)} \left(\begin{array}{c}
\delta \bbox{r}^{(n)} \\
\delta \bbox{p}^{(n)}
\end{array}\right)\,.
\end{equation}
Here ${\bf F}^{(n)}$ is a matrix with elements
\begin{equation}
F_{\mu\nu}^{(n)} =m \gamma \int_{n\delta t}^{(n\!+\!1)\delta t}\!\!
{\rm d}t' \,\,\frac{\partial u_\mu(\bbox{r}_{t'},t')}{\partial r_\nu}
\end{equation}
with $\mu,\nu = 1,2,3$, and ${\bf I}$ is the $3\times 3$ unit
matrix. The random components $F_{\mu\nu}^{(n)}$ average to zero,
are uncorrelated for $n\neq n'$, and the correlation $\langle
F_{\mu\nu}^{(n)} F_{\mu'\nu'}^{(n)}\rangle$ is determined by the
statistical properties of $\bbox{u}(\bbox{r},t)$ described in
section \ref{sec: 3.1}. The Lyapunov exponents are identified as
the asymptotic growth rates  of the three largest eigenvalues  of
the product ${\bf M}^{(N)} {\bf M}^{(N-1)}\cdots {\bf M}^{(0)}$
nearly diagonal matrices for large values of $N$. The asymptotic
growth rates are determined using a method described in
\cite{Eck85}.

\section{Langevin equations for Lyapunov exponents}
\label{sec: 3}

\subsection{Equations in Langevin form}
\label{sec: 3.1}

The equation (\ref{eq: 2.25}) for ${\bf R}'$ can be simplified
when the correlation time of the velocity field is sufficiently
short, and when the amplitude of the random force is sufficiently
small. In this limit the force-gradient term ${\bf F}'$ behaves like a
white-noise signal, and the equations of motion reduce to a
system of Langevin equations:
\begin{equation}
\label{eq: 3.1} {\rm d}{\bf R}'=\biggl(-\gamma {\bf R}'
 -\frac{1}{m} {{\bf R}'}^2 +[{\bf R}',{\bf O}^+ \dot{\bf O}]\biggr)
 {\rm d}t
+{\rm d}\bbox{\zeta}
\end{equation}
where ${\rm d}\bbox{\zeta}$ is a matrix of
random increments ${\rm d}\zeta_{\mu\nu}$ satisfying
\begin{equation}
\label{eq: 3.2}
\langle {\rm d}\zeta_{\mu\nu}\rangle=0
\ ,\ \ \
\langle {\rm d}\zeta_{\mu\nu}{\rm d}\zeta_{\mu'\nu'}\rangle=
2{\cal D}_{\mu\nu,\mu'\nu'}{\rm d}t
\ .
\end{equation}
The elements ${\cal D}_{\mu\nu,\mu'\nu'}$ of the \lq diffusion
matrix' depend on the statistics of the fluid velocity field
$\mbox{\boldmath$u$}(\mbox{\boldmath$r$},t)$. The latter may be
decomposed into potential and solenoidal components, generated
from scalar and vector potentials. The Stokes drag force on the
particle, $\mbox{\boldmath$f$}=m\gamma \mbox{\boldmath$u$}$ is
written in terms of potentials $\phi$ and $\mbox{\boldmath$A$}$:
\begin{equation}
\label{eq: 3.5} \mbox{\boldmath$f$}=\mbox{\boldmath$\nabla$}\phi
+\mbox{\boldmath$\nabla$} \wedge \mbox{\boldmath$A$}\ .
\end{equation}
The fields $\phi(\mbox{\boldmath$r$},t)\equiv A_0(\mbox{\boldmath$r$},t)$ and
$A_\mu(\mbox{\boldmath$r$},t),\,\mu=1,2,3$ are assumed to possess statistical
properties which are homogeneous in space and time, and
isotropic in space. Also, it is assumed that these fields
are uncorrelated, and the intensity of the $A_\mu$ fields
(for $\mu=1,2,3$) is such that the correlation function is of
the form
\begin{equation}
\label{eq: 3.6} \langle
A_\mu(\mbox{\boldmath$r$}_0+\mbox{\boldmath$R$},t_0+t)
A_\nu(\mbox{\boldmath$r$}_0,t_0)\rangle
=\delta_{\mu\nu}[(1-\alpha^2)\delta_{\mu 0}+\alpha^2] \,C(R,t)
\end{equation}
for some constant $\alpha$, where $R=\vert
\mbox{\boldmath$R$}\vert$.

Using spatial and temporal homogeneity of the velocity field, and
noting that in the limit ${\rm Ku}\to 0$ the particle does not
move significantly in time $\tau$, elements of the diffusion
matrix are
\begin{equation}
\label{eq: 3.7}
{\cal D}_{\mu\nu,\mu'\nu'}={\textstyle{1\over 2}}\int_{-\infty}^\infty
{\rm d}t\ \biggl\langle{\partial f_\mu\over{\partial r_\nu}}({\bbox{0}},t)
{\partial f_{\mu'}\over{\partial r_{\nu'}}}(\bbox{0},0)
\biggr\rangle
\ .
\end{equation}
Inserting the expression (\ref{eq: 3.5}) for the force
gives an expression for these elements in terms of second derivatives of
the fields $A_\mu$. For any isotropic field $A(\mbox{\boldmath$r$},t)$,
these satisfy, for example,
\begin{equation}
\label{eq: 3.8}
\biggl\langle {\partial^2 A\over{\partial r_1^2}}(\mbox{\boldmath$r$},t)
{\partial^2 A\over{\partial r_1^2}}(\mbox{\boldmath$r$},0)\biggl\rangle
=3\biggl\langle {\partial^2 A\over{\partial r_1\partial r_2}}
(\mbox{\boldmath$r$},t)
{\partial^2 A\over{\partial r_1\partial r_2}}(\mbox{\boldmath$r$},0)
\biggl\rangle
=3\biggl\langle {\partial^2 A\over{\partial r_1^2}}(\mbox{\boldmath$r$},t)
{\partial^2 A\over{\partial r_2^2}}(\mbox{\boldmath$r$},0)\biggl\rangle
\ .
\end{equation}
Now consider the evaluation of the elements of the diffusion matrix,
${\cal D}_{\mu\nu,\mu'\nu'}$. First we note that due to isotropy, the values
of the elements are invariant under any permutation of indices
(for example, ${\cal D}_{33,22}={\cal D}_{11,33}$). It is easy to see that
${\cal D}_{\mu\nu,\mu'\nu'}=0$ unless the four indices can be grouped into
two equal pairs. There are four cases where indices are paired,
namely ${\cal D}_{aa,bb}$, ${\cal D}_{ab,ab}$, ${\cal D}_{ab,ba}$
and ${\cal D}_{aa,aa}$
where $a$ and $b$ are different numbers from the set $\{1,2,3\}$.
We define
\begin{equation}
\label{eq: 3.9} {\cal D}_0={\textstyle{1\over
2}}\int_{-\infty}^\infty {\rm d}t\ \biggl\langle {\partial^2
\phi\over{\partial r_1^2}}({\bf 0},t) {\partial^2 \phi
\over{\partial r_1^2}}({\bf 0},0)\biggr\rangle
\end{equation}
and use (\ref{eq: 3.8}) to express the non-zero elements
${\cal D}_{\mu\nu,\mu'\nu'}$ in terms of ${\cal D}_0$ and $\alpha^2$. It
is simplest to calculate specific examples of the non-zero
elements, and to deduce the others by permuting indices:
writing $\partial^2 A_\mu/\partial r_\nu\partial r_\rho(\mbox{\boldmath$r$},t)$
in shorthand as $(\partial^2_{\nu\rho} A_\mu)_t$, we have
\begin{eqnarray}
\label{eq: 3.10} {\cal D}_{11,11}&=&{\textstyle{1\over
2}}\int_{-\infty}^\infty {\rm d}t\ \langle
(\partial^2_{11}\phi+\partial^2_{21}A_3-\partial^2_{31}A_2)_t
(\partial^2_{11}\phi+\partial^2_{21}A_3-\partial^2_{31}A_2)_0\rangle
\nonumber \\
&=&{\cal D}_0\biggl(1+{2\alpha^2\over 3}\biggr)\equiv {\cal D}_1
\nonumber \\
{\cal D}_{12,12}&=&{\textstyle{1\over 2}}\int_{-\infty}^\infty
{\rm d}t\ \langle
(\partial^2_{12}\phi+\partial^2_{22}A_3-\partial^2_{32}A_2)_t
(\partial^2_{12}\phi+\partial^2_{22}A_3-\partial^2_{32}A_2)_0\rangle
\nonumber \\
&=&{\cal D}_0\biggl({1\over 3}+{4\alpha^2\over 3}\biggr)\equiv {\cal D}_2
\nonumber \\
{\cal D}_{11,22}&=&{\textstyle{1\over 2}}\int_{-\infty}^\infty
{\rm d}t\ \langle (\partial^2_{11}
\phi+\partial^2_{21}A_3-\partial^2_{31}A_2)_t
(\partial^2_{22}\phi+\partial^2_{32}A_1-\partial^2_{21}A_3)_0\rangle
\nonumber \\
&=&{\cal D}_0\biggl({1\over 3}-{\alpha^2\over 3}\biggr)\equiv {\cal D}_3
\nonumber \\
{\cal D}_{12,21}&=&{\textstyle{1\over 2}}\int_{-\infty}^\infty
{\rm d}t\ \langle
(\partial^2_{12}\phi+\partial^2_{22}A_3-\partial^2_{32}A_2)_t
(\partial^2_{12}\phi
+\partial^2_{31}A_1-\partial^2_{11}A_3)_0\rangle
\nonumber \\
&=&{\cal D}_0\biggl({1\over 3}-{\alpha^2\over 3}\biggr)={\cal D}_3
\end{eqnarray}
where the first three relations define the constants ${\cal D}_1$,
${\cal D}_2$ and ${\cal D}_3$.

The Langevin equations (\ref{eq: 3.1})
can be labelled by a single index, which will
be indicated by using a Roman letter, and packing the double indices
such that $i=3(\mu-1)+\nu$. The above Langevin equations are then
of the form
\begin{equation}
\label{eq: 3.3} {\rm d}R_i'=\biggl[-\gamma R_i'{\rm
d}t-{1\over{m}}\sum_{j=1}^9\sum_{k=1}^9 V_{ijk} R_j'
R_k'\biggr]{\rm d}t+{\rm d}\zeta_i
\end{equation}
with $\langle {\rm d}\zeta_i{\rm d}\zeta_j\rangle=2{\cal
D}_{ij}{\rm d}t$. The $V_{ijk}$ are determined by (\ref{eq: 3.1});
most of them are zero. It is convenient to scale the equations for
$R'_i$ to dimensionless form. We write
\begin{equation}
\label{eq: 3.12} t'=\gamma t \ ,\ \ \ x_i=\sqrt{{\gamma\over{{\cal
D}_1}}}R_i \ ,\ \ \ {\rm d}w_i=\sqrt{{\gamma\over{{\cal
D}_1}}}{\rm d}\zeta_i
\end{equation}
and
\begin{equation}
\label{eq: 3.13} \epsilon={{\cal D}_1^{1/2}\over{m\gamma^{3/2}}} \
,\ \ \ \Gamma={1+4\alpha^2\over{3+2\alpha^2}} \ .
\end{equation}
The parameter $\epsilon$ is a dimensionless measure of the
particle inertia, and the parameter $\Gamma$ characterises the
nature of the flow: it ranges between ${\textstyle{1\over 3}}$ and
$2$, and we have
\begin{equation}
\label{eq:gamma}
\Gamma
= \left\{\begin{array}{ll}
{\textstyle{1\over 3}} & \mbox{for potential flow}\\
2 & \mbox{for solenoidal flow\ .}
\end{array}\right .
\end{equation}
In \cite{Meh04} where two-dimensional flows are discussed,
the corresponding parameter satisfies $1/3\leq \Gamma \leq 3$,
and solenoidal flow is obtained for $\Gamma = 3$.

In scaled coordinates, the Langevin equations are
\begin{equation}
\label{eq: 3.14}
{\rm d}x_i=-\biggl[x_i+\epsilon \sum_{j=1}^9\sum_{k=1}^9 V_{ijk}
x_j x_k\biggr]{\rm d}t'+{\rm d}w_i \equiv v_i {\rm d}t' + {\rm d}w_i
\end{equation}
where a \lq velocity' $\bbox{v}$ with components
\begin{eqnarray}
\label{eq:v}
&&v_i =  v_i^{(0)} + \epsilon  v_i^{(1)} \\
\label{eq:v2}
&&v_i^{(0)} = -x_i\,,\quad v_i^{(1)}
= -\sum_{j=1}^9\sum_{k=1}^9 V_{ijk} x_j x_k
\end{eqnarray}
was introduced, and
$\langle {\rm d}w_i{\rm d}w_j\rangle=2D_{ij}{\rm d}t'$.
The transformed diffusion matrix ${\bf D}$ with elements
$D_{ij}$ is given by
\begin{equation}
\label{eq: 3.15}
{\bf D}=\pmatrix{
 1 & 0 & 0 & 0 &\sigma& 0 & 0 & 0 &\sigma\cr
 0 &\Gamma& 0 &\sigma& 0 & 0 & 0 & 0 & 0 \cr
 0 & 0 &\Gamma& 0 & 0 & 0 &\sigma& 0 & 0 \cr
 0 &\sigma& 0 &\Gamma& 0 & 0 & 0 & 0 & 0 \cr
\sigma& 0 & 0 & 0 & 1 & 0 & 0 & 0 &\sigma\cr
 0 & 0 & 0 & 0 & 0 &\Gamma& 0 &\sigma& 0 \cr
 0 & 0 &\sigma& 0 & 0 & 0 &\Gamma& 0 & 0 \cr
 0 & 0 & 0 & 0 & 0 &\sigma& 0 &\Gamma& 0 \cr
\sigma& 0 & 0 & 0 &\sigma& 0 & 0 & 0 & 1 }\,\ \ \
\sigma={\textstyle{1\over 2}}(1-\Gamma)\,.
\end{equation}
Note that $\Gamma={\cal D}_2/{\cal D}_1$ and $\sigma={\cal
D}_3/{\cal D}_1$. The diffusion matrix can be set in block-diagonal
form, with three $2\times 2$ blocks and one $3\times 3$
block, which is in cyclic form: it can therefore be diagonalised
analytically.

\subsection{Discussion}
\label{sec: 3.2}

The problem of determining the Lyapunov exponent of particles
suspended in a turbulent fluid has thus been transformed into the
problem of determining expectation values of the position of a
particle undergoing a Langevin process in a nine-dimensional space
(defined by equation (\ref{eq: 3.14})). From (\ref{eq: 3.12}) we
see that $\langle R'_i\rangle/m=\gamma \epsilon \langle
x_i\rangle$, and from (\ref{eq: 2.20}) we see that the Lyapunov
exponents are
\begin{equation}
\label{eq: 3.16}
\lambda_1=\gamma \epsilon \langle x_1\rangle
\ ,\ \ \
\lambda_2=\gamma \epsilon \langle x_5\rangle
\ ,\ \ \
\lambda_3=\gamma \epsilon \langle x_9\rangle
\ .
\end{equation}
The velocity (\ref{eq:v})  in the Langevin equation contains terms which are
linear in the displacement, $\bbox{v}^{(0)}$, which drive this Langevin particle
back towards the origin. If, however, the particle diffuses
sufficiently far from the origin, the quadratic terms $\bbox{v}^{(1)}$  in the
velocity become dominant, and may take the particle away to
infinity. In fact, numerical studies show that the particle always
does escape to infinity. We must consider the significance of this
effect. When the inertia of the suspended particles is
large, the momenta of the particles are not determined
solely by their positions. It is therefore possible for two of the
vectors $\delta \mbox{\boldmath$r$}_\mu$ to become coplanar, while
the vectors $\delta \mbox{\boldmath$p$}_\mu$ continue to span
three dimensions. As the vectors $\delta \mbox{\boldmath$r$}_\mu$
approach co-planarity, the inverse of the matrix ${\bf R}$ defined
by (\ref{eq: 2.6}) must become singular. Correspondingly, some or
all of the components $x_i$ must diverge to infinity in a finite
time. After the point $\mbox{\boldmath$x$}(t)$ diverges to
infinity, continuity of $\delta \mbox{\boldmath$r$}_\mu$ and
$\delta \mbox{\boldmath$p$}_\mu$ implies that it immediately
reappears and converges from the reflected point at infinity. In
terms of the parametrisation of the displacements $\delta
\mbox{\boldmath$r$}_\mu$ given in (\ref{eq: 2.3}), we see that two
of the vectors become collinear when $\delta\phi=0$ and all three
are collinear when $\delta\theta=0$. The $\delta \phi=0$ is
therefore a co-dimension one-condition, which is realised by
varying time. When $\delta\phi=0$ the elements of the second
and third columns of ${\bf R}'$ diverge.
Generically, the condition $\delta \theta=0$ never
occurs.

The events where the elements of ${\bf R}'$ diverge in the
Langevin simulation therefore correspond to events where the local
density of the suspended particles diverges because the volume
element of their flow vanishes \cite{Wil05}. These events are
termed caustics. Simulation of the Langevin equations therefore
also gives the rate $J$ at which an particle passes through a
caustic (which is identical to the rate at which Langevin
trajectories escape to infinity), as well as an estimate of the
Lyapunov exponents.

The Langevin equations (\ref{eq: 3.1}) are a valid approximation
for (\ref{eq: 2.25}) and (\ref{eq: 2.26}) provided that two
conditions are satisfied. The correlation time of the forcing
terms must be small compared to the relaxation time $\gamma^{-1}$:
this clearly implies that the Stokes number should be large, that
is $\Omega \ll 1$. A second condition is that the random forcing
term should be sufficiently weak that the displacement of the
coefficients $R'_{\mu \nu}$ during the correlation time $\tau $
should be small, relative to their typical values. This condition
was discussed in \cite{Meh05}: in the notation of this present
paper it leads to the requirement that $\kappa \ll 1$. The
arguments of this section are therefore valid in the limit where
$\Omega \ll 1$ and $\kappa \ll 1$.

Figure \ref{fig:f2} illustrates the application of the Langevin
equations (\ref{eq: 3.14}) to calculating the Lyapunov exponents
for a flow with a very short correlation time. The results are
compared with a direct evaluation obtained by multiplying the
monodromy matrices of the corresponding flow, as explained
in section 3.2. The first three
panels show the Lyapunov exponents as a function of $\epsilon$ for
three different values of $\Gamma$, corresponding to
incompressible flow ($\Gamma=2$), purely potential flow
($\Gamma=\frac{1}{3}$), and a mixed case, $\Gamma=1$. The final
panel illustrates the limiting behaviour of the Lyapunov exponents
as $\epsilon \to \infty$. In that limit the effect of the damping
$\gamma$ becomes negligible, and the Lyapunov exponents reach
limiting values that are independent of $\gamma $. This implies
that the functions $f_j(\epsilon)=\lambda_j/\gamma$ are asymptotic
to
\begin{equation}
\label{eq: 3.17} f_j(\epsilon)\equiv \frac{\lambda_j}{\gamma}\sim
C_j\epsilon^{2/3}
\end{equation}
in the limit as $\epsilon \to \infty$, for some coefficients $C_j$
which depend upon $\Gamma$. We have not been able to determine the
coefficients $C_j(\Gamma)$ analytically.

The Langevin equation does not appear to be exactly solvable.
Section \ref{sec: 10} considers how to obtain a perturbation
series solution for the variables $\langle x_i\rangle$.

\subsection{Estimate for diffusion constant in turbulent flow}
\label{sec: 3.3}

In section \ref{sec: 3.1} we discussed the definition of
$\epsilon$ for a flow with small Kubo number. However, the
principal area of application of our results is to particles
suspended in a fully developed turbulent flow. In section
\ref{sec: 1.2} we argued that the Kubo number is always of order
unity for this case. Here we consider how the definition of
$\epsilon $ must be adapted to make our results applicable to
turbulent flows, and what is the appropriate value for the
correlation time $\tau$.

The dimensionless parameter $\epsilon$ is expressed (equation
(\ref{eq: 3.13})) in terms of a diffusion constant defined by
(\ref{eq: 3.9}). This quantity is related to the spectral
intensity of the rate of strain in the neighbourhood of a particle
with trajectory $\mbox{\boldmath$r$}(t)$, defined by
\begin{equation}
\label{eq: 3.18} I(\omega)=\int_{-\infty}^\infty {\rm d}t\
\exp({\rm i}\omega t)\bigg\langle \frac{\partial u_1}{\partial
x_1}(\bbox{r}(t),t) \frac{\partial u_1}{\partial
x_1}(\bbox{0},0)\bigg\rangle\ .
\end{equation}
Note that (unlike (\ref{eq: 3.9})) we consider the temporal
variation of the position $\mbox{\boldmath$r$}(t)$ of the
suspended particle, because when ${\rm Ku}=O(1)$ this may change
by a significant amount during the correlation time, $\tau$. Let
us first consider how to estimate this quantity when the Stokes
number of the suspended particles is small, so that the
correlation function in (\ref{eq: 3.18}) may be approximated by a
Lagrangian correlation function. We estimate ${\cal D}\propto
m^2\gamma^2I(0)$, so that $\epsilon^2\propto I(0)/\gamma$. The
intensity $I(0)$ has the dimension of inverse time, so that
$I(0)\propto \tau_{\rm s}^{-1}$, where $\tau_{\rm s}$ is a
characteristic time scale of the flow. In the case of fully
developed turbulence, it is not immediately clear whether
$\tau_{\rm s}$ is the time scale associated with the dissipation
scale, or else some longer time scale. The Kolmogorov theory of
turbulence \cite{Fri97} implies that if the integral defining
$I(\omega)$ is dominated by the inertial range, we may write
$I(\omega)$ in terms of ${\cal E}$, the rate of dissipation per
unit mass, with no dependence upon $\nu$. We therefore seek a
relation of the form
\begin{equation}
\label{eq: 3.19} I(\omega)=C {\cal E}^\alpha \omega^\beta
\end{equation}
for some constants $\alpha$, $\beta$, $C$. This relation is only
dimensionally consistent for $\alpha=0$, $\beta=1$, which makes
the result independent upon ${\cal E}$ (and furthermore implies
that it vanishes at $\omega=0$). We infer that the value of ${\cal
D}_0$ is not determined by the inertial range of turbulent flow,
and we expect that it is determined by the dissipative scale, with
characteristic time scale $\tau\sim \sqrt{\nu/{\cal E}}$. We
therefore expect that
\begin{equation}
\label{eq: 3.20} \epsilon^2={K\over{\gamma}} \sqrt{\frac{\cal
E}{\nu}}
\end{equation}
for fully-developed turbulence, where Kolmogorov's 1941 theory of
turbulence \cite{Fri97} predicts that $K$ is a universal constant.

Finally we consider two refinements of the estimate (\ref{eq:
3.20}). Recent insights concerning intermittency suggest
\cite{Fri97} that $K$ should have a weak dependence upon Reynolds
number of the turbulence. More significantly, at very large values
of the Stokes number, equation (\ref{eq: 3.20}) may overestimate
$\epsilon^2$ since $\partial_{x_1} u_1(\bbox{r}(t),t)$ will
fluctuate more rapidly than it would for a particle which is
advected. (An analogous effect is described in detail in
\cite{Arv06}.) These considerations suggest that we write
\begin{equation}
\label{eq: 3.21} \epsilon^2={K({\rm St},{\rm Re})\over{\gamma}}
\sqrt{\frac{\cal E}{\nu}}
\end{equation}
where ${\rm St}=\sqrt{{\cal E}/\nu\gamma^2}$ is the Stokes number
referred to the Kolmogorov timescale. The function $K({\rm
St},{\rm Re})$ approaches a finite limit as ${\rm St}\to 0$, but
approaches zero as ${\rm St}\to \infty$, and has a weak dependence
upon the Reynolds number, ${\rm Re}$.

\section{Perturbation theory}
\label{sec: 10}

\subsection{Mapping to Hamiltonian form}
\label{sec: 10.1}

The Langevin equations (\ref{eq: 3.14}) are equivalent to a
Fokker-Planck equation for the probability density
$P(\bbox{x},t')$ of $\mbox{\boldmath$x$}$
\begin{equation}
\label{eq: 10.0}
\label{eq:FP}
{\partial P \over{\partial t'}}
=\bbox{\nabla}\cdot \left(-\bbox{v}  + {\bf D}\bbox{\nabla}\right)P
\end{equation}
When $\epsilon=0$, the velocity $\bbox{v}$ of the Fokker-Planck
equation (\ref{eq:FP}) is linear in the displacement from the
origin. see equations (\ref{eq:v},\ref{eq:v2}). It is easily
verified that the solution is a Gaussian function. This suggests
that it may be possible to map the unperturbed ($\epsilon=0$)
problem to a nine-dimensional harmonic oscillator. We shall show
how this mapping can be achieved, and what is the form of the
perturbation representing the velocity terms which are quadratic
functions of the coordinates.

We write the Fokker-Planck equation as
\begin{eqnarray}
\label{eq: 10.1}
{\partial P \over{\partial t'}}
&\equiv&\hat{F}P= (\hat{F}_0 + \epsilon \hat{F}_1)P
\end{eqnarray}
where the notation $\hat A$ indicates that the object
$A$ is an operator. We have
\begin{equation}
\label{eq: 10.2} \hat{F}_0=\bbox{\nabla}\cdot\left(-\bbox{v}^{(0)}
+ {\bf D}\bbox{\nabla}\right) \ ,\ \ \
\hat{F}_1=-\bbox{\nabla}\cdot\bbox{v}^{(1)}
\end{equation}
with $\bbox{v}^{(0)}=-\bbox{x}$ and where the components
of $\bbox{v}^{(1)}$ are the quadratic terms in $\bbox{v}$
proportional to $\epsilon$, defined in (\ref{eq:v2}).
We consider the steady-state solution which solves
$\hat{F}P(\mbox{\boldmath$x$})= 0$
for $\epsilon=0$: this is
\begin{equation}
\label{eq: 10.3} P_0(\bbox{x})=A \exp(-{\textstyle{1\over
2}}\bbox{x}\cdot {\bf D}^{-1}\,\bbox{x}) \equiv A
\exp[-\Phi_0(\bbox{x})]
\end{equation}
(where $A$ is a normalisation constant). The latter identity
defines $\Phi_0(\bbox{x})$. We now define $\hat H$ by
\begin{equation}
\label{eq: 10.4}
\hat{H}=\exp(\Phi_0/2) \hat{F} \exp(-\Phi_0/2)
\end{equation}
and re-write the steady-state Fokker-Planck equation as
\begin{equation}
\label{eq: 10.5}
\hat{H}Q(\bbox{x}) = 0
\end{equation}
where $Q(\bbox{x})=\exp[\Phi_0(\bbox{x})/2]P(\bbox{x})$. This form
will be suitable for a perturbation expansion in the small
parameter $\epsilon$, since $\hat{H}$ consists of two parts, i.e.,
$\hat{H}=\hat{H}_0 + \epsilon \hat{H}_1$, with $\hat H_0$ a
Hermitian operator.

We require an explicit expression for
$\hat{H}_0=\exp(\Phi_0/2) \hat{F}_0 \exp(-\Phi_0/2)$.
By considering the action of $\hat{H}_0$ upon an arbitrary function $f$
we find
\begin{eqnarray}
\label{eq: 10.6}
\hat{H}_0&=&{\textstyle{1\over 2}}
\sum_i \delta_{ii}
-{\textstyle{1\over 4}}\sum_{ij} x_i (D^{-1})_{ij} x_j
+\sum_{ij} D_{ij}\, \partial^2_{ij}
\nonumber \\
&=&{\textstyle{1\over 2}}{\rm tr}({\bf I})
-{\textstyle{1\over 4}}\bbox{x} \cdot {\bf D}^{-1} \, \bbox{x}
+\partial_{\bbox{\scriptstyle x}} \cdot {\bf D}\,
\partial_{\bbox{\scriptstyle x}}
\end{eqnarray}
where ${\bf I}$ is the $9\times 9$ unit matrix and
$\partial_{\bbox{\scriptstyle x}} \equiv
(\partial_{x_1},\ldots,\partial_{x_9})\equiv
(\partial_1,\ldots,\partial_9)$. Note that (\ref{eq: 10.6}) is the
Hamiltonian operator of an isotropic nine-dimensional quantum
harmonic oscillator.

Next consider $\hat{H}_1=\exp(\Phi_0/2)\hat{F}_1 \exp(-\Phi_0/2)$:
again we consider the action of $\hat{H}_1$ upon an arbitrary function $f$
\begin{eqnarray}
\label{eq: 10.7}
-\hat{H}_1 f
&=&\sum_{i}\exp(\Phi_0/2)\,\partial_i(v_i^{(1)}\exp(-\Phi_0/2) f)
\nonumber \\
&=&\sum_i \partial_i (v_i^{(1)} f)-{\textstyle{1\over 2}}
\sum_{ij} (D^{-1})_{ij} x_j v_i^{(1)} f
\end{eqnarray}
so that
\begin{equation}
\label{eq: 10.8} \hat{H}_1=-\partial_{\bbox{\scriptstyle x}} \cdot
\bbox{v}^{(1)} +{\textstyle{1\over 2}}\bbox{v}^{(1)}\cdot{\bf
D}^{-1}  \bbox{x} \ .
\end{equation}

\subsection{Use of annihilation and creation operators}
\label{sec: 10.2}

We next consider how to re-write the operators $\hat{H}_0$ and
$\hat{H}_1$ in terms of harmonic-oscillator creation an
annihilation operators. This is easier if we first diagonalise
$\hat{H}_0$. The diffusion matrix $\bf D$ is a real symmetric
matrix, which can be diagonalised by an orthogonal matrix, ${\bf
U}$
\begin{equation}
\label{eq: 10.9} {\bf U}^{-1} {\bf D} {\bf U}=\bbox{\Lambda}
\end{equation}
where $\bbox{ \Lambda}$ is a diagonal matrix with elements
$\Lambda_{ij}=\Lambda_i \delta_{ij}$, and $\Lambda_i$ are the
eigenvalues of ${\bf D}$. We find
\begin{equation}
\label{eq: 10.10} \hat{H}_0={\textstyle{1\over 2}}{\rm tr}({\bf
I})- {\textstyle{1\over 4}}{\bbox{\eta}} \cdot
\bbox{\Lambda}^{-1}\, \bbox{\eta} +\partial_{\bbox{\scriptstyle
\eta}} \cdot \bbox{\Lambda}\,
\partial_{\bbox{\scriptstyle \eta}}
\end{equation}
where
\begin{equation}
\label{eq: 10.11} \bbox{\eta} = {\bf U}^{-1}\,\bbox{x} \ ,\ \ \
\partial_{\bbox{\scriptstyle \eta}}
= {\bf U}^{-1}\, \partial_{\bbox{\scriptstyle x}} \ .
\end{equation}
Since $\bbox{\Lambda}$ is diagonal, equation (\ref{eq: 10.10})
simplifies to
\begin{equation}
\label{eq: 10.12} \hat{H}_0= {\textstyle{9\over 2}}
+{\textstyle{1\over 2}}\sum_i\biggl(2\Lambda_i
\frac{\partial^2}{\partial \eta_i^2} -\frac{1}{2\Lambda_i}
\eta_i^2\biggr) \ .
\end{equation}
Now define the following operators
\begin{eqnarray}
\label{eq: 10.13} \hat p_i &=& -{\rm i}\, \sqrt{2\Lambda_i}\,
\frac{\partial}{\partial \eta_i}
\nonumber \\
\hat q_i &=& \frac{1}{\sqrt{2\Lambda_i}}\, \eta_i
\end{eqnarray}
which are analogous to the position and momentum operators of
quantum theory. In terms of $q_i$ and $p_i$ (\ref{eq: 10.12})
becomes
\begin{equation}
\label{eq: 1.14}
\hat{H}_0=-{\textstyle{1\over 2}}\sum_i(\hat p_i^2 + \hat q_i^2)+
{\textstyle{9\over 2}}\hat I
\end{equation}
where $\hat I$ is the identity operator.
This is the Hamiltonian of an isotropic quantum harmonic oscillator
in nine dimensions.
The commutator of the $\hat q_i$ and $\hat p_i$ operators is
$[\hat q_i,\hat p_j]={\rm i}\hat I\delta_{ij}$,
so that the pairs $\hat q_i$ and $\hat p_i$ have the same commutators as the
position and momentum operators in quantum mechanics (with $\hbar=1$).
We next introduce
\begin{eqnarray}
\label{eq: 10.15}
\hat a_i &=& \frac{\hat q_i +{\rm i} \ \hat p_i }{ \sqrt{2} }
\nonumber \\
\hat a_i^+ &=& \frac{\hat q_i -{\rm i} \ \hat p_i }{ \sqrt{2} }
\end{eqnarray}
whose commutator is $[\hat a_i,\hat a_j^+]=\delta_{ij}\hat I$.
The $\hat a^+_i$ and $\hat a_i$ are respectively the
creation and annihilation operators, or raising and lowering
operators, for the degree of freedom labelled by $i$. Now we find
\begin{equation}
\label{eq: 10.16}
\hat{H}_0=-\sum_{i=1}^{9}  \hat a^+_i \hat a_i
\ .
\end{equation}
Reducing the unperturbed operator to this form is useful because
the simple algebraic properties of the annihilation and creation
operators make it possible to perform perturbation theory exactly,
to any desired order. This is achieved by using the set of
eigenfunctions of $\hat H_0$ as a convenient basis set. These
eigenfunctions are labelled by a set of quantum numbers
$\{m_1,m_2,..,m_9\}$, taking values $m_j=0,1,2,3...$. Thus each
eigenfunction is labelled by a vector $\mbox{\boldmath$m$}$ with
non-negative integer elements. The eigenfunction with this set of
quantum numbers will be denoted by a Dirac \lq ket' vector, $\vert
\mbox{\boldmath$m$})$ \cite{Dir30}. (We use this notation rather
than the more usual $\vert \mbox{\boldmath$m$}\rangle$ to avoid
confusion with the angular brackets denoting averages.) For the
Hamiltonian (\ref{eq: 10.16}), the eigenvalues are just the sum of
the quantum numbers, so that the eigenvalue equation is written
\begin{equation}
\label{eq: 10.16a} \hat H_0 \vert \mbox{\boldmath$m$}
)=-\sum_{j=1}^9 m_j \vert \mbox{\boldmath$m$})\ .
\end{equation}
The annihilation $\hat a_j$ and creation $\hat a^+_j$ operators
map one eigenstate to another, by (respectively) raising and lowering
the quantum number $m_j$:
\begin{eqnarray}
\label{eq: 10.16b} \hat a_j \, \vert m_1,m_2,..,m_j,..,m_9)
&=&\sqrt{\,m_j\hspace*{-2mm}\phantom{1}}\,\,\vert m_1,m_2,..,m_j-1,..,m_9)
\nonumber \\
\hat a_j^+ \, \vert m_1,m_2,..,m_j,..,m_9) &=&\sqrt{m_j+1}\,\, \vert
m_1,m_2,..,m_j+1,..,m_9)
\end{eqnarray}
and if $\hat a_j$ acts on an eigenstate for which the quantum
number $m_j$ is zero, the result is zero.

We now consider how to rewrite $\hat{H}_1$ in terms of the $\hat
a_i$ and $\hat a^+_i$ operators following the same procedure as
for $\hat{H}_0$. Using (\ref{eq: 10.11}) we have
\begin{equation}
\label{eq: 10.17}
\partial_{\bbox{\scriptstyle x}} \cdot \bbox{v}^{(1)}
=\sum_{ij} U_{ji} \partial_{\eta_i} v^{(1)}_j \ ,\ \ \
\bbox{v}^{(1)} \cdot {\bf D}^{-1}\,  \bbox{x} =\sum_{ij} v^{(1)}_j
U_{ji} \ \Lambda_i^{-1} \eta_i \ .
\end{equation}
We obtain
\begin{equation}
\label{eq: 10.19} \hat{H}_1= -\sum_{ij} U_{ji}
\left(\frac{\partial}{\partial \eta_i} -{\textstyle{1\over
2}}\Lambda_i^{-1} \eta_i\right)
v^{(1)}_j\big(\bbox{x}(\bbox{\eta})\big)
\end{equation}
where we have made the dependence of the components of
$\bbox{v}^{(1)}(\bbox{x}(\bbox{\eta}))$ upon the variables
$\bbox{\eta}$ explicit.

The final step is to express $\hat{H}_1$ in terms of the $\hat
a_i$ and $\hat a_i^+$ operators. Inserting $\eta_i = (\hat a_i + \hat
a_i^+) \sqrt{\Lambda_i}$ and $\partial/\partial \eta_i=(\hat a_i
-\hat a_i^+)/ 2\sqrt\Lambda_i$ into equation (\ref{eq: 10.19}) yields
\begin{equation}
\label{eq: 10.21} \hat{H}_1 =\sum_{ij} U_{ji}
\frac{1}{\sqrt{\Lambda_i}} \ \hat a^+_i
v^{(1)}_j(\bbox{x}(\bbox{\eta})) \ .
\end{equation}
Now each of the $v_j^{(1)}$ can be expressed as a quadratic form
\begin{equation}
\label{eq: 10.22} v_j^{(1)}(\bbox{x}) =-\sum_{kl} V_{jkl} \,x_k x_l
\end{equation}
where $V_{jkl}$ are the coefficients (almost all zero) of the
terms $x_k x_l$ appearing in each $v_j^{(1)}$.
Since $\bbox{x}(\bbox{\eta})={\bf U} \bbox{\eta}$ and $\eta_i$ are
given by rearranging (\ref{eq: 10.15}), we have
\begin{equation}
\label{eq: 10.23} x_k=\sum_m U_{km}(\hat a_m + \hat a_m^+)
\sqrt{\Lambda_m}\ .
\end{equation}
Combining with (\ref{eq: 10.22}) and inserting into (\ref{eq:
10.21}) gives
\begin{equation}
\label{eq: 10.24}
\hat{H}_1=\sum_{imn} H^{(1)}_{imn}\,
\hat a^+_i(\hat a_m+\hat a_m^+)(\hat a_n+\hat a_n^+)
\end{equation}
where
\begin{equation}
\label{eq: 10.25} H^{(1)}_{imn}=-
\sqrt{{\Lambda_m\Lambda_n\over{\Lambda_i}}} \sum_{jkl}
V_{jkl}U_{ji}U_{km}U_{ln} \ .
\end{equation}
Note that the coefficients $H_{imn}^{(1)}$ defining the
perturbation operator $\hat H_1$ can be obtained exactly, because
the matrix ${\bf D}$ can be diagonalised exactly.

\section{Iterative development of the perturbation series}
\label{sec: 6}

\subsection{Method for developing the series expansion}

Instead of solving the Fokker-Planck equation $\hat F\,
P(\mbox{\boldmath$x$})=0$ we attempt to solve $\hat
H\,Q(\mbox{\boldmath$x$})=0$, where $Q(\mbox{\boldmath$x$})
=\exp[\Phi_0(\mbox{\boldmath$x$})/2]P(\mbox{\boldmath$x$})$. In
the following we use a shorthand notation for integrals which is
equivalent to the Dirac notation \cite{Dir30} of quantum
mechanics: given two functions $a(\mbox{\boldmath$x$})$ and
$b(\mbox{\boldmath$x$})$ and an operator $\hat A$, we define
\begin{equation}
\label{eq: 6.0} (a\vert \hat A\vert b)\equiv \int {\rm
d}\mbox{\boldmath$x$}\ a^\ast(\mbox{\boldmath$x$})\,\hat A\,
b(\mbox{\boldmath$x$})\equiv \int_{-\infty}^\infty {\rm d}x_1
\ldots \int_{-\infty}^\infty {\rm d}x_9\ a^\ast(x_1,\ldots ,x_9)\,
\hat A\, b(x_1,\ldots,x_9)\ .
\end{equation}
Here the asterisk denotes complex conjugation.
Now consider how to obtain the Lyapunov exponents from
the function $Q$. They are obtained from the mean values
$\langle x_i\rangle$ of the coordinates. These are
\begin{eqnarray}
\label{eq: 6.1}
\langle x_i\rangle&=&\int {\rm d}\mbox{\boldmath$x$}\
x_i\, P(\mbox{\boldmath$x$})
\nonumber \\
&=&\frac{\int {\rm d}\mbox{\boldmath$x$}\
\exp[-\Phi_0(\mbox{\boldmath$x$})/2]
\,x_i\,Q(\mbox{\boldmath$x$})}{\int {\rm
d}\mbox{\boldmath$x$}\exp[-\Phi_0(\mbox{\boldmath$x$})]Q(\mbox{\boldmath$x$})}
\nonumber \\
&=&\frac{({\bf 0}\vert \hat x_i\vert Q)}{({\bf 0}\vert
Q)}\nonumber \\
&=&{1\over{({\bf 0}\vert Q)}} \sum_j U_{ij}\sqrt{\Lambda_j}({\bf
0}\vert \hat a_j+\hat a^+_j\vert Q)
\end{eqnarray}
where we have used the fact that
$\exp[-\Phi_0(\mbox{\boldmath$x$})/2]$ is the eigenfunction of the
\lq ground state', $\vert {\bf 0})$. The denominator is included
to take account of normalisation.

We calculate $\vert Q)$ by perturbation theory: writing
\begin{equation}
\label{eq: 6.2}
\vert Q)=\vert Q_0)+\epsilon \vert Q_1)+\epsilon^2\vert Q_2)+...
\end{equation}
we find that the functions $\vert Q_k)$ satisfy the recursion relation
\begin{equation}
\label{eq: 6.3}
\vert Q_{k+1})=-\hat H_0^{-1}\hat H_1 \vert Q_k)
\ .
\end{equation}
At first sight this appears to be ill-defined because one of the
eigenvalues of $\hat H_0$ is zero, so that the inverse of $\hat
H_0$ is only defined for the subspace of states which are
orthogonal to the \lq ground state', $\vert {\bf 0})$. However,
because all of the components of $\hat H_1$ have a creation
operator as a left factor, the state $\hat H_1\vert \psi)$ is
orthogonal to $\vert {\bf 0})$ for any function $\vert \psi )$, so
that (\ref{eq: 6.3}) is in fact well-defined. The iteration starts
with $\vert Q_0)=\vert {\bbox{0}})$.

The functions $\vert Q_k)$
are expanded in terms of harmonic-oscillator eigenstates
$\vert \bbox{m})$,
with coefficients $a_{\bbox{\scriptstyle m}}^{(k)}$  and
$\bbox{m} = (m_1,\ldots,m_9)$:
\begin{equation}
\label{eq: 6.4}
\vert Q_k)=\sum_{\bbox{\scriptstyle m}}
a^{(k)}_{\bbox{\scriptstyle m}}
\vert \bbox{m})
\ .
\end{equation}
By projecting equation (\ref{eq: 6.4}) onto the vector
$\vert\bbox{m})$ and using the fact that the eigenvectors
$\vert \bbox{m}')$ of $\hat H_0$ form a complete
basis, the iteration
can be expressed as follows (for $\bbox{m}\ne \bbox{0}$):
\begin{equation}
\label{eq: 6.5}
a^{(k+1)}_{\bbox{\scriptstyle m}} =
\sum_{\bbox{\scriptstyle m}'}
\frac{(\bbox{m}\vert \hat H_1\vert \bbox{m}')}{\sum_j m_j}
a^{(k)}_{\bbox{\scriptstyle m}'}
\ .
\end{equation}
The matrix elements $(\bbox{m}\vert\hat H_1\vert\bbox{m}')$ are
readily computed using the algebraic properties of the raising and
lowering operators, (\ref{eq: 10.16b}) \cite{Dir30}. The
coefficients $a^{(k)}_{\bbox{\scriptstyle m}}$ are then calculated
recursively, so that we obtain the states $|Q_k)$. Finally,
expectation values are extracted using (\ref{eq: 6.1}).

\subsection{Programming the perturbation expansion}
\label{sec: 6.2}

It is not practicable to carry out the perturbation expansion by
hand even at leading order, because the Hamiltonian $\hat H_1$
contains several hundred non-zero coefficients $H^{(1)}_{ijk}$.
The calculation was automated using a {\it Mathematica} program.

From (\ref{eq: 6.3}) we see that the $n^{\rm th}$ order of the
perturbation expansion requires the calculation of \lq matrix
elements' of the form
\begin{equation}
\label{eq: 13.0} I(n,j)=({\bf 0}\vert (\hat a_j+\hat a^+_j)\hat
H_0^{-1}\hat H_1 \hat H_0^{-1} \hat H_1 \ldots \hat H_0^{-1}\hat
H_1\vert {\bf 0})
\end{equation}
with $\hat H_1$ occurring $n$ times. Because the Hamiltonian has
been expressed in terms of raising and lowering operators, each
successive application of $\hat H_1$ or $\hat H_0^{-1}$ to the
eigenfunction $\vert {\bf 0})$ gives a function which consists of
a linear combination of a finite number of eigenfunctions $\vert
\mbox{\boldmath$m$})$. The inner product is then readily evaluated
using the orthonormality of the eigenfunctions:
\begin{equation}
\label{eq: 13.1} (\mbox{\boldmath$m$}\vert
\mbox{\boldmath$m$}')=\delta_{\bbox{\scriptstyle m},\bbox{\scriptstyle m}'}
=\prod_{i=1}^9
\delta_{m_i,m'_i}\ .
\end{equation}
The number of eigenfunctions included in the sum (\ref{eq: 6.4})
increases very rapidly as the order of the perturbation increases.
The matrix element $I(n,j)$ can be written as the inner product of
two function vectors:
\begin{eqnarray}
\label{eq: 13.2} I(n,j)&=&(\chi(k,j)\vert \psi(n-k))\nonumber\\
\vert \chi(k,j))&=&\underbrace{\hat H_1\hat H_0^{-1}\hat H_1\hat
H_0^{-1}\ldots \hat H_1\hat H_0^{-1}(\hat a_j+\hat a_j^+)\vert
{\bf 0})}_{\hat H_1 {\rm appears\
}k{\rm \ times}}\nonumber \\
\vert \psi(m))&=&\underbrace{\hat H_0^{-1}\hat H_1\ldots \hat
H_0^{-1}\hat H_1\vert {\bf 0})}_{\hat H_1{\rm \ appears\ }m{\rm \
times}}
\end{eqnarray}
where the choice of $k$ is, in principle, arbitrary. However, the
computational effort is proportional to the sum of the number of
terms in the vectors $\vert \chi(n,k))$ and $\vert \psi(m))$, and
this is minimised by taking $k\approx n/2$ in equation (\ref{eq:
13.2}).

The program was checked by comparing the coefficients for
$\lambda_1$ with those determined in \cite{Meh05} using
a perturbation of a two-dimensional harmonic oscillator, which
allowed evaluation of the first $47$ non-vanishing coefficients. We also
remark that for the second Lyapunov exponent it is sufficient to
consider a system of six coupled harmonic oscillators, because the
Langevin equations for $(x_1,\ldots,x_6)$ do not depend upon the
values of $(x_7,x_8,x_9)$. This allows the perturbation
series to be taken to higher order for the second Lyapunov
exponent.

\subsection{Results}

We find that the denominator $({\bf 0}\vert Q)$ in (\ref{eq: 6.1})
is equal to unity to all orders in $\epsilon$, and from the
numerator we obtain series expansions for the Lyapunov exponents
in the form
\begin{eqnarray}
\label{eq: 6.6} \lambda_j/\gamma  &=& \sum_{n=1}^\infty
c^{(j)}_n(\Gamma)\, \epsilon^{2n} \ .
\end{eqnarray}
Note that only even orders in $\epsilon $ occur in this expansion:
this fact is most readily understood using an argument which will
be given in section \ref{sec: WKB}. The first seven coefficients
$c_n^{(j)}$ are
\begin{eqnarray}
\label{eq: 6.7}
c_1^{(1)}(\Gamma) &=& -1+2\Gamma       \nonumber  \\
c_2^{(1)}(\Gamma) &=& -5+20\, \Gamma-16\,\Gamma^2           \nonumber\\
c_3^{(1)}(\Gamma) &=& -60+360\,\Gamma-568\,\Gamma^2+272\,\Gamma^3           \nonumber\\
c_4^{(1)}(\Gamma) &=& -1105 + 8840\,\Gamma -61936\,\Gamma^2/3+58432\,\Gamma^3/3 -19648\,\Gamma^4/3
\nonumber\\
c_5^{(1)}(\Gamma)
&=&-27120+271200\,\Gamma-7507040\,\Gamma^2/9+3492160\,\Gamma^3/3\nonumber\\
&&\hspace*{5mm}-2316032\,\Gamma^4/3+ 1785856 \,\Gamma^5/9 \nonumber\\
c_6^{(1)}(\Gamma) &=& -828250 + 9939000\,\Gamma - 1020068800\,\Gamma^2/27 + 1874157440\,\Gamma^3/27\nonumber\\
&&\hspace*{5mm}- 613664384\,\Gamma^4/9 + 934756352\,\Gamma^5/27 - 193558528\,\Gamma^6/27\nonumber\\
c_7^{(1)}(\Gamma)
&=& -30220800 + 423091200\,\Gamma
- 154727293760\,\Gamma^2/81 + 351319669760\,\Gamma^3/81 \nonumber\\
&&\hspace*{5mm}
- 454943581760\,\Gamma^4/81
+ 342675611776\,\Gamma^5/81 - 140392616704^6/81 \nonumber\\
&&\hspace*{5mm}+ 24271797760\,\Gamma^7/81 \nonumber\\
&& \nonumber\\[3mm]
c_1^{(2)}(\Gamma) &=& -2+\Gamma \nonumber        \\
c_2^{(2)}(\Gamma) &=& -16 + 22\,\Gamma - 5\,\Gamma^2           \nonumber\\
c_3^{(2)}(\Gamma) &=& -549/2 +1287\, \Gamma/2 - 871\,\Gamma^2/2 + 125\, \Gamma^3/2           \nonumber\\
c_4^{(2)}(\Gamma) &=& -13463/2 + 22506\,\Gamma - 79121\,\Gamma^2/3 + 35606\,\Gamma^3/3 - 7723\,\Gamma^4/6
\label{eq: 6.8}\\
c_5^{(2)}(\Gamma)
&=& -627719/3 + 2731795\,\Gamma/3 - 13765330\,\Gamma^2/9 + 3598630\,\Gamma^3/3\nonumber\\
&&\hspace*{5mm}- 1229363\,\Gamma^4/3 + 341381\,\Gamma^5/9\nonumber\\
c_6^{(2)}(\Gamma)
&=& -280669811/36 + 250852811\,\Gamma/6 - 9891631295\,\Gamma^2/108 + 2783144725\,\Gamma^3/27\nonumber\\
&&\hspace*{5mm}- 2202644629\,\Gamma^4/36 + 924868595\,\Gamma^5/54
- 157226321\,\Gamma^6/108
    \nonumber\\
c_7^{(2)}(\Gamma)
&=&
-145680639449/432 +
       928376776943\,\Gamma/432 - 7524732877927\,\Gamma^2/1296\nonumber\\
      &&\hspace*{5mm} + 11046167913985\,\Gamma^3/1296
              - 9388185321985\,\Gamma^4/1296 +
                     4517559789671\,\Gamma^5/1296 \nonumber\\
                     &&\hspace*{5mm}-
                     1087174658765\,\Gamma^6/1296 +
                     87859310987\,\Gamma^7/1296
    \nonumber\\
&&\nonumber\\[3mm]
c_1^{(3)}(\Gamma) &=& -3 \nonumber       \\
c_2^{(3)}(\Gamma) &=& -33 + 12\,\Gamma           \nonumber\\
c_3^{(3)}(\Gamma) &=& -1479/2 + 1215\,\Gamma/2 - 429\,\Gamma^2/2 - 3\,\Gamma^3/2          \nonumber\\
c_4^{(3)}(\Gamma) &=& -45627/2 + 29954\,\Gamma - 21071\,\Gamma^2 + 5394\,\Gamma^3 + 143\,\Gamma^4/2
\nonumber\\
c_5^{(3)}(\Gamma)
&=& -1731931/2 + 18888175\,\Gamma/12 - 4963475\,\Gamma^2/3 + 5130835\,\Gamma^3/6 \nonumber\\
&&\hspace*{5mm}- 1013585\,\Gamma^4/6 - 37765\,\Gamma^5/12 \nonumber\\
c_6^{(3)}(\Gamma) &=& -461447213/12 + 268463707\,\Gamma/3 - 2249372585\,\Gamma^2/18 + 872716325\,\Gamma^3/9 \nonumber\\
&&\hspace*{5mm}- 1414393915\,\Gamma^4/36 + 56257766\,\Gamma^5/9 + 1316179\,\Gamma^6/9 \nonumber\\
c_7^{(3)}(\Gamma) &=&
-279693223943/144 +
       1190601381865\,\Gamma/216 - 4142788537873,\Gamma^2/432
       \nonumber\\
 &&\hspace*{5mm}      + 2141088699035\,\Gamma^3/216
 - 290883665975\,\Gamma^4/48 + 144871714325\,\Gamma^5/72\nonumber\\
&&\hspace*{5mm} - 38215846457\,\Gamma^6/144 - 1592440259\,\Gamma^7/216
\nonumber\ .\\
&&\nonumber
\end{eqnarray}
Equations (\ref{eq: 6.6}) and  (\ref{eq: 6.7}) are the main results
of this paper. They provide an expansion of the Lyapunov exponents
in terms of the dimensionless parameter $\epsilon \sim \kappa
\Omega^{-1/2}$. The expansion is valid when $\kappa \ll 1$ (small
Kubo number) and $\Omega \ll 1$ (large Stokes number; this is the
underdamped limit where inertial effects are dominant).

The coefficients of $\Gamma$ in (\ref{eq: 6.7}) are all rational
numbers, despite the fact that the algebra of the raising and
lowering operators (equations (\ref{eq: 10.16b})) produces
expressions containing square roots of integers. The reason for
cancellation of the square roots to produce rational coefficients
is not yet understood.

In the remaining sections we  discuss physical and mathematical
implications of  (\ref{eq: 6.6}) and  (\ref{eq: 6.7}), and relate
these results to results obtained in the overdamped limit (where
inertial effects are small).

\subsection{Summation of the perturbation series}

\begin{figure}
\centerline{\includegraphics[width=6cm]{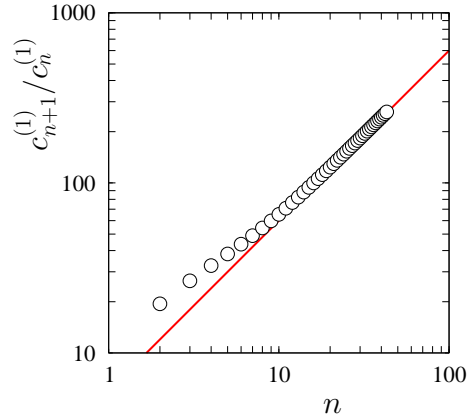}}
\caption{\label{fig:f3} Illustrating the growth of the
perturbation coefficients $c^{(1)}_n$ with increasing order $n$
for the first Lyapunov exponent, $\Gamma = 2$. Shown is the ratio
$c^{(1)}_{n+1}/c^{(1)}_{n}$ as a function of $n$. The solid line
is the asymptotic relation $c^{(1)}_{n+1}/c^{(1)}_n=6\,n$,
consistent with $c^{(1)}_{n} \sim C\, 6^n\, (n-1)!$, for some
constant $C$.}
\end{figure}

The coefficients in (\ref{eq: 6.8}) grow rapidly with order $n$,
indicating that these are divergent series. We find they have
factorial growth for large $n$:
\begin{equation}
\label{eq:growth} c_n^{(j)} \sim C {S^n}\,(n-1)!
\end{equation}
for some constant $C$. This factorial growth is typical of an
asymptotic series \cite{Din74}. We find that $S =1/6$, independent
of $j$ and $\Gamma$. Figure ~\ref{fig:f3} shows the growth of
$c_n^{(1)}$ for $\Gamma = 2$.

There are rather few concrete physical problems for which
perturbation expansion coefficients are available to high order:
most studies are concerned with quantum-mechanical problems which
are perturbations of a harmonic oscillator, such as \cite{Gra70}.
Here we do have the perturbation series coefficients to high
order, for the same underlying reason: we used the algebra of
harmonic-oscillator raising and lowering operators to compute
matrix elements exactly.

Methods for treating divergent series are discussed in \cite
{Din74} and \cite{Boy99}, assuming that the expansion can be
pursued to high order. Here we use Pad\'e-Borel summation, similar
to an approach used to sum the perturbation series of certain
one-dimensional quantum-mechanical anharmonic oscillators
\cite{Gra70}. We evaluate the \lq Borel sum'
\begin{equation}
\label{eq:borelsum} B_j(\epsilon^2) = \sum_{n=1}^{n_{\rm max}}
\frac{c_n^{(j)}}{n!} \epsilon^{2n}
\end{equation}
where $n_{\rm max}$ is the number of terms available. The sum of
the series is then taken to be
\begin{equation}
\label{eq:boreltransform} \lambda_j/\gamma = \mbox{Re}\int_{\cal
C} {\rm d}t\,  B_j(\epsilon^2 \,t) {\rm e}^{-t}
\end{equation}
where ${\cal C}$ is a suitable curve in the complex plane.
Assuming that the Borel sum has a finite radius of convergence, the function
may be approximated by Pad\'{e} approximants of order $N/M$,
$P_{N/M}(\epsilon^2 t)$ \cite{Ben78}.  Here $N$ and $M$ are the orders of
the numerator and denominator polynomials, respectively and
$N+M \leq n_{\rm max}$, where $n_{\rm max}$ is the number
of coefficients available.

The integral in (\ref{eq:boreltransform}) is evaluated
numerically. If the Pad\'{e} approximants to $B_j(\epsilon^2 t)$
have poles on the positive real axis, the integration path in
(\ref{eq:boreltransform}) is taken to be a ray in the upper right
quadrant in the complex plane.

\begin{figure}
\centerline{\includegraphics[width=12cm]{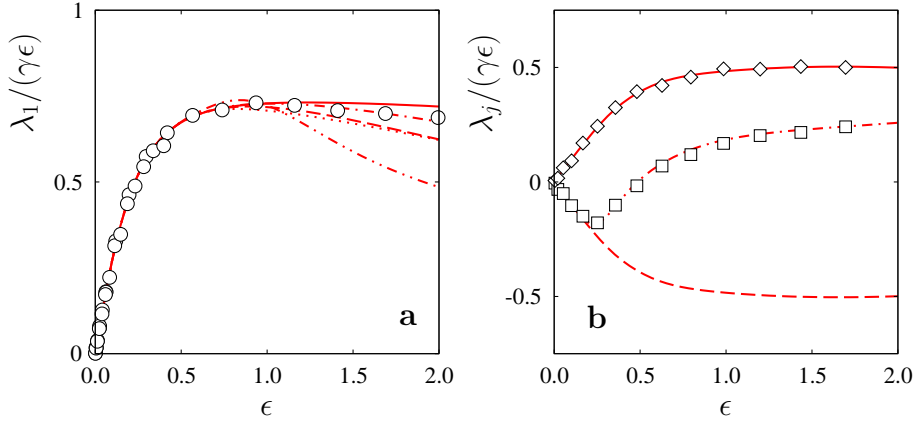}}
\caption{\label{fig:f4} Borel-Pad\'e{} summation of the
perturbation expansion of $\lambda_j$. {\bf a}: Numerical data for
$\lambda_1$ as a function of $\epsilon$ for $\Gamma=2$ ($\circ$)
(obtained as described in section \ref{sec:numcalc}), compared
with results of Borel-Pad\' e summation, equations
(\ref{eq:borelsum}) and (\ref{eq:boreltransform}); $P_{5/5}$ ($- -
-$), $P_{10/10}$ ($-\cdot - \cdot -$), $P_{15/15}$ ($-\cdot\cdot -
\cdot\cdot- $), and $P_{20/20}$
($\cdot\cdot\cdot\cdot\cdot\cdot\cdot$). and  $P_{23/23}$
($-\!\!\!-\!\!\!-\!\!\!-\!\!\!-\!\!\!-$). {\bf b}: Numerical data
for $\lambda_1$ ($\diamond$) and $\lambda_2$ ($\Box$) as a
function of $\epsilon$ for $\Gamma=1$. Also shown are results of
Pad\' e-Borel summation (using a Pad\'e{} approximant $P_{23/23}$)
for $j=1$ (solid line) and $j=2$ (dashed line). Since according to
equation (\ref{eq: 6.7}) $c_l^{(2)} = -c_l^{(1)}$ at $\Gamma=1$,
those curves are identical apart from an overall minus sign. But
using the relation $\lambda_1+\lambda_2\sim
C\exp[-1/(6\,\epsilon^2)]$, with $\lambda_1$ obtained by
Borel-Pad\'e summation and $C=0.79$ a fitted parameter ($-\cdot -
\cdot -$) gives a much better fit to $\lambda_2$ than the dashed
line.}
\end{figure}

Results of Pad\'e-Borel summation of the series for $\lambda_1$ in
the incompressible case ($\Gamma=2$) are shown in figure
\ref{fig:f4}{\bf a} for $N=M=5,10,15,20,23$. The summation results
are in good agreement with numerical results for $\lambda_1$
obtained as described in section \ref{sec:numcalc}. For the first
Lyapunov exponent, we obtained higher-order coefficients using the
method described in \cite{Meh05} (which is particularly efficient
but also restricted to calculating the maximal exponent): in this
case $n_{\rm max}= 47$. We conclude that the Pad\'e-Borel approach
works very well for the largest Lyapunov exponent.

Figure \ref{fig:f4}{\bf b} shows results for the first and second
Lyapunov exponents for $\Gamma=1$, representing a flow field with
both solenoidal and compressible components.  For the maximal
Lyapunov exponent, $\lambda_1$, Pad\'e-Borel summation gives very
good agreement with the numerical results. For the second Lyapunov
exponent, inspection of the coefficients (\ref{eq: 6.7}) shows
that for $\Gamma=1$, $c_l^{(2)} = -c_l^{(1)}$, in other words, the
perturbation coefficients for $\lambda_1+\lambda_2$ vanish
identically for $\Gamma=1$. However direct numerical simulations
show that $\lambda_1+\lambda_2$ is not equal to zero when
$\Gamma=1$. The WKB analysis summarised in section \ref{sec: WKB}
gives rise to the hypothesis that there may be a non-analytical
contribution to the Lyapunov exponents of the form
$C_j(\Gamma,\epsilon) \exp(-1/6\epsilon^2)$ which has to be added
to the Pad\'{e}-Borel approximation. Figure \ref{fig:f4}{\bf b}
shows that this is indeed the case for the second Lyapunov
exponent for $\Gamma=1$: adding $C\,\exp(-1/6\epsilon^2)$ (with
$C=0.79$ a fitted constant) to the Pad\'e-Borel sum for
$\lambda_2$ (solid line) gives very good agreement with the
numerical data. This observation shows that there are
contributions to the Lyapunov exponents which are not captured by
a simple application of the Borel-Pad\'e summation technique.

Figure \ref{fig:f5} shows a comparison of the results of
Borel-Pad\'e summation with Monte-Carlo simulations of all three
of the three Lyapunov exponents for $\Gamma =2$ (incompressible
flow). Here we used the coefficients given in section \ref{sec:
6.2} with $n_{\rm max} = 7$. The results for the first Lyapunov
exponent show excellent agreement, as mentioned above. Those for
the second exponent show a systematic deviation for larger values
of $\epsilon$. The results for the third exponent show some
instability upon changing the number of terms included in the
Borel sum, implying that more terms are required to achieve a
convergent result.

We conclude that the Borel-Pad\'e summation
gives very satisfactory results for small
values of $\epsilon$, but for large $\epsilon$ there are
systematic deviations which are not yet understood. These appear
to be associated with non-analytic contributions of the form
$\exp(-\Phi^\ast/\epsilon^2)$.

\subsection{Relation to clustering}

A major motivation for calculating the Lyapunov exponents is to
evaluate the dimension deficit $\Delta$ (defined by equation
(\ref{eq: 0.0})) and hence the Lyapunov dimension, $d_{\rm
L}=3-\Delta$. To lowest order in $\epsilon$, equations (\ref{eq:
6.6}) and (\ref{eq: 6.7}) imply that
\begin{equation}
\label{eq:underdamped}
\lambda_1/\gamma = (-1+2\Gamma)\epsilon^2\,,\hspace*{5mm}
\lambda_2/\gamma = (-2+\Gamma)\epsilon^2\,,\hspace*{5mm}
\lambda_3/\gamma = -3\,\epsilon^2
\ .
\end{equation}
In figure \ref{fig:f2} {\bf a}, {\bf b}, and {\bf c} these
expressions are shown as dashed lines. Equations
(\ref{eq:underdamped}) imply that in for an incompressible flow
($\Gamma=2$), the sum $\lambda_1+\lambda_2+\lambda_3$ vanishes in
the limit of $\epsilon \rightarrow 0$, and so does $\Delta$,
defined in (\ref{eq: 0.0}). This reflects that in the absence of
inertial effects, an incompressible flow cannot give rise to
density fluctuations. The leading-order behaviour of the dimension
deficit is
\begin{eqnarray}
\label{eq:sum} \Delta &=& \frac{3\Gamma-6}{2\Gamma-1}
+\frac{54-54\,\Gamma+21\,\Gamma^2}{2 \Gamma-1}\,\epsilon^2
+O(\epsilon^4)\ .
\end{eqnarray}
In the incompressible
case this gives $\Delta = 10\,\epsilon^2$ for small $\epsilon$.

Figure \ref{fig:f1}{\bf a} shows numerical results for $\Delta $
as a function of $\epsilon$ for incompressible flow, including the
asymptotic approximation (\ref{eq:sum}). When the numerical
results were re-plotted in figure \ref{fig:f1}{\bf b}, we used the
relation ${\rm St}=\epsilon^2/0.256^2$, with the factor chosen to
give a good agreement as judged by eye.

\begin{figure}
\centerline{\includegraphics[width=12cm]{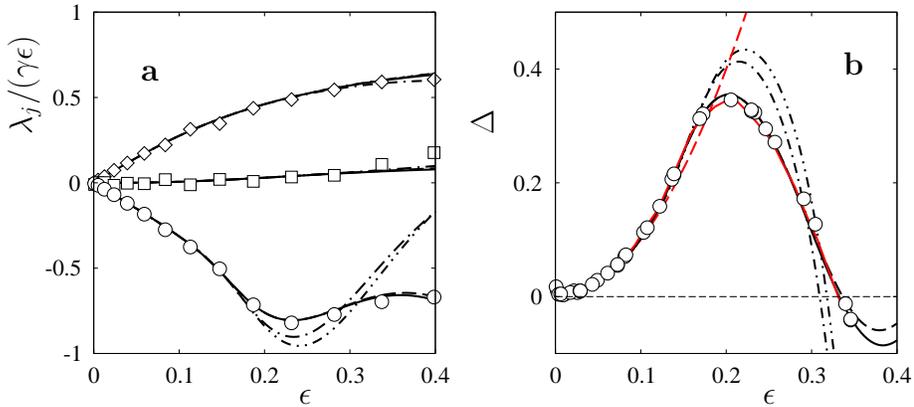}}
\caption{\label{fig:f5} {\bf a} dependence of the Lyapunov
exponents on $\epsilon$, for $\Gamma=2$: the symbols show
numerical data obtained as described in section \ref{sec:numcalc},
and the solid lines are the results of Borel-Pad\' e summation.
The results for $\lambda_3$ depend upon the number of orders of
the perturbation series included in the Borel sum, $P_{2/2}$
($-\cdot\cdot - \cdot\cdot-$), $P_{2/3}$ ($-\cdot - \cdot-$),
$P_{3/3}$ ($- - - $), and $P_{3/4}$
($-\!\!\!-\!\!\!-\!\!\!-\!\!\!-\!\!\!-$). {\bf b} dimension
deficit obtained from the curves in figure \ref{fig:f5}{\bf a}
using equation (\ref{eq: 0.0}): the same line styles are used for
different Pad\'e approximants. The solid red curve shows the
results of Monte-Carlo simulations of the Langevin equation
(\ref{eq: 3.14} - \ref{eq: 3.16}), and is the same as in figure
\ref{fig:f1}{\bf a}. The dashed red line is the asymptotic
approximation (\ref{eq:sum}).}
\end{figure}

Figure \ref{fig:f5}{\bf b} shows results for the dimension deficit
$\Delta$ obtained from Pad\'e-Borel summations for the Lyapunov
exponents shown in figure \ref{fig:f5}{\bf a}, using equation
(\ref{eq: 0.0}). For small values of $\epsilon$ the agreement is
excellent. Despite shortcomings of the Borel summations for
$\lambda_2$ and $\lambda_3$ mentioned above, we observe
satisfactory agreement between the theory and results of computer
simulations for all values of  $\epsilon$ where $\Delta$ is
positive. In summary, Pad\'e{}-Borel summation of the series
(\ref{eq: 6.8}) provides a satisfactory theoretical description of
the dimension deficit $\Delta$. The maximal value of $\Delta$
obtained here (and in \cite{Dun05}) is in good agreement with
direct numerical simulations of particles suspended in a
Navier-Stokes flow (see figure \ref{fig:f1}).

\section{WKB analysis}
\label{sec: WKB}

In this section we discuss the stationary solution of the
Fokker-Planck equation (\ref{eq:FP}) by means of a WKB method,
along the lines discussed in \cite{Fre84}. This gives an
indication as to the interpretation of the non-analytic
contributions to the Lyapunov exponents of the form
$\exp(-\Phi/\epsilon^2)$ which were considered in section
\ref{sec: 6}. A precise evaluation of these contributions is
beyond the scope of existing methods.

It is convenient to re-scale the variables $x_i$ in (\ref{eq:FP})
using $x_i' = x_i \,\epsilon$. The stationary Fokker-Planck
equation for $P(\bbox{x}')$ is then
\begin{equation}
\label{eq: 12.2} \nabla'\cdot\bigl[(-\bbox{v}'+\epsilon^2{\bf
D}\bbox{\nabla}') P(\bbox{x}')\bigr]
 = 0
\end{equation}
with $v_i' = -x_i' -\sum_{jk} V_{ijk} x_j' x_k'$. Equation
(\ref{eq: 12.2}) indicates why the odd orders of the perturbation
series developed in section \ref{sec: 6} vanish: because this
equation contains $\epsilon^2$ rather than $\epsilon$, expansion
of any quantity as a power series must be a series in powers of
$\epsilon^2$.

In the remainder of this section we drop the primes and make the
WKB ansatz \cite{Ben78}
\begin{equation}
\label{eq: 12.3} P(\bbox{x}) = A
\exp\left[-\frac{1}{\epsilon^2}\left( \Phi(\bbox{x}) +
\epsilon^2\Psi(\bbox{x}) + \cdots\right)\right]
\end{equation}
(where $A$ is a normalisation factor). Our objective is to
determine the nature of the function $\Phi$ in (\ref{eq: 12.3}).
Inserting this into (\ref{eq: 12.2}) and keeping only the
lowest-order terms in $\epsilon$ gives
\begin{equation}
\label{eq: 12.4} \bbox{v}\cdot \bbox{\nabla} \Phi +
\bbox{\nabla}\Phi \cdot{\bf D} \bbox{\nabla}\Phi = 0 \ .
\end{equation}
This has the form of a Hamilton-Jacobi equation \cite{Fre84}.
Recall that given a Hamiltonian $H(\bbox{x},\bbox{p})$ the
Hamilton-Jacobi equation for the action $\Phi(\bbox{x},E)$ is
\begin{equation}
\label{eq: 12.5}
 H(\bbox{x},\bbox{\nabla}\Phi) =  E \ .
\end{equation}
The Hamilton-Jacobi equation is solved by finding the classical
trajectories which are solutions of Hamilton's equations of
motion. The function $\Phi(\bbox{x},E)$ is then obtained by
integration along the classical trajectory
\begin{equation}
\label{eq: 12.6} \Phi(\bbox{x},E) =
\int_0^{t(\bbox{\scriptstyle x},\scriptstyle E)}\!\!\!{\rm d}t\ \dot{\bbox{x}}\cdot \bbox{p}
\ .
\end{equation}
In our case, $E=0$ and the Hamiltonian is
\begin{equation}
\label{eq: 12.7} H(\bbox{x},\bbox{p}) = \bbox{v}(\bbox{x})\cdot
\bbox{p} +\bbox{p}\cdot {\bf D}\,\bbox{p}
\end{equation}
and (using the fact that ${\bf D}$ is a symmetric matrix)
Hamilton's equations of motion are
\begin{equation}
\label{eq: 12.8} \dot{\bbox{x}} = \frac{\partial H}{\partial
\bbox{p}} = \bbox{v}(\bbox{x}) + 2{\bf D}\,\bbox{p} \ ,\ \ \
\dot{\bbox{p}} = -\frac{\partial H}{\partial \bbox{x}} = -{\bf
F}\, \bbox{p}
\end{equation}
where the matrix ${\bf F}$ has components $F_{ij} = \partial
v_i/\partial x_j$. Close to the origin, $\Phi(\bbox{x})$ is
expected to be approximately
$\Phi_0(\bbox{x})=\frac{1}{2}\bbox{x}\cdot\bf{D}\,\bbox{x}$, in
agreement with the Gaussian approximate solution (\ref{eq: 10.3}).
The appropriate initial conditions for the trajectories are
therefore infinitesimally close to the origin in
$\bbox{x}$-$\bbox{p}$-space with initial condition $\bbox{p} =
\bbox{\nabla} \Phi(\bbox{x})=\bbox{x}$.

Singular points of $\Phi(\bbox{x})$ are points $\bbox{x}^\ast$
where $\bbox{\nabla}\Phi = \bbox{0}$. Consider a trajectory from
the origin passing through a singular point $\bbox{x}^\ast$. At
this point $\bbox{\nabla}\Phi$ vanishes and consequently also
$\bbox{p}$. From this point onwards, $\dot{\bbox{p}}=\bbox{p}=0$
and after reaching the singular point the dynamics is thus
advective, $\dot{\bbox{x}} = \bbox{v}(\bbox{x})$, with $\Phi$
remaining constant at $\Phi^\ast = \Phi(\bbox{x}^\ast)$. Singular
points are expected to give rise to non-analytic behaviour of the
Lyapunov exponents of the form $\exp(-\Phi^\ast/\epsilon^2)$.

There is a trajectory of equations (\ref{eq: 12.8}) for which all
of the coordinates except $x_1$ are zero, for which there is a
singular point at $x_1^\ast=-1$ with action
$\Phi^\ast=\Phi(x_1^\ast)=1/6$. We can therefore expect
non-analytic contributions to the Lyapunov exponents of the form
$F\exp(-1/6\epsilon^2)$, where $F$ may have an algebraic dependence
upon $\epsilon$.

\begin{figure}
\centerline{\includegraphics[width=6cm]{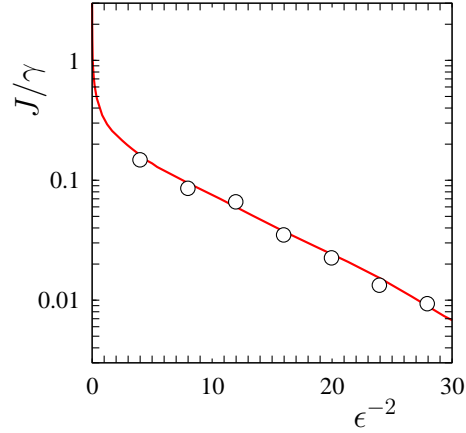}}
\caption{\label{fig:f6} Rate of caustic formation $J'= J/\gamma$
as a function of $\epsilon$, for $\Gamma=2$, Monte-Carlo
simulations of the Langevin equation (\ref{eq: 3.14} - \ref{eq:
3.16}), solid red line, and numerical data obtained as described
in section \ref{sec:numcalc} ($\circ$). The results are asymptotic
to $J'=C\exp(-\Phi_{\rm caustic}/\epsilon^2)$ in the limit as
$\epsilon \to 0$, with $C \approx 0.28$ and $\Phi_{\rm
caustic}\approx 0.125$ (dashed line). }
\end{figure}

The formation of caustics is associated with escape of the
Langevin trajectory to infinity. The Langevin equations have an
attractive fixed point at $\mbox{\boldmath$x$}={\bf 0}$, but
particles that diffuse sufficiently far from this fixed point will
escape to infinity. We expect that the escape rate contains a
factor $\exp[-\Phi^\ast/\epsilon^2]$, where $\Phi^\ast$ is the
action of a trajectory to a singular point. We find that the rate
of caustic formation is of the form
\begin{equation}
\label{eq: 12.9} J'\sim J'_0\exp(-\Phi^\ast_{\rm
caustic}/\epsilon^2)
\end{equation}
Numerical results confirming this expectation are illustrated in
figure \ref{fig:f6}, which shows that the action associated with
the formation of caustics is $\Phi^\ast_{\rm caustic}\approx
0.125$.

\section{Advective approximation}
\label{sec: 7}

In the remaining two sections we discuss the overdamped limit
$\Omega\gg 1$ and consider how this connects with our results for
the underdamped limit described in sections 3 to 7. A surprising
fact is that the leading-order formulae for the expansion of the
Lyapunov exponents in $\kappa$ are identical in the limits
$\Omega\ll 1$ and $\Omega\gg 1$ (although the higher-order terms
differ). When $\Omega \gg 1$ the particles are advected by the
flow. However, we show that when $\Omega \ll 1$, despite the fact
that the particles are not advected, the equations determining
their separation are identical to those of advected particles in
the limit $\epsilon \to 0$. It is this advective approximation
which is discussed in this section.

In the advective  limit it is of interest to compute  the sum
of the three Lyapunov exponents for particles suspended in an
incompressible fluid beyond the leading-order approximation. This
case is important because its sign determines whether particles in
an incompressible flow cluster. The result for purely advected
particles is zero, so that a theory going beyond the advective
approximation is required. This special case is considered in
section \ref{sec: 8}.

This section is structured as follows. First in section \ref{sec:
7.1} we discuss the advective approximation and the conditions for
its validity. We also show that the Lyapunov exponents can be
obtained from an It\^ o type Brownian process. The Lyapunov
exponents of this process were calculated by Le Jan \cite{LeJ85},
using a notation and terminology which are difficult to relate to
our own. In section \ref{sec: 7.2} we calculate the Lyapunov
exponents using a simpler method, expressed in our own notation.
We calculate the leading-order expressions for the Lyapunov
exponents for the model with force statistics defined by equation
(\ref{eq: 3.6}), considered as an expansion in $\kappa$, showing
that these are identical when $\Omega\gg 1$ to those expressions
obtained in section \ref{sec: 6} when $\Omega\ll 1$.

\subsection{Brownian advection model}
\label{sec: 7.1}

Consider the linearised equations of motion.
These can be integrated to give
\begin{equation}
\label{eq: 11.1}
\delta \mbox{\boldmath$r$}(\Delta t)-\delta \mbox{\boldmath$r$}(0)
={1\over{m}}\int_0^{\Delta t}{\rm d}t_1\ \delta \mbox{\boldmath$p$}(t_1)
\end{equation}
and
\begin{equation}
\label{eq: 11.2}
\delta \mbox{\boldmath$p$}(t_1)=\int_{-\infty}^{t_1} {\rm d}t_2\
\exp[-\gamma (t_1-t_2)]\,{\bf F}(\mbox{\boldmath$r$}(t_2),t_2)\,
\delta \mbox{\boldmath$r$}(t_2)
\ .
\end{equation}
We now consider the mean and variance of the displacement
(\ref{eq: 11.1}).

\subsubsection*{Mean value of displacement}

Combining these expressions, writing the result in component
form, and expanding the functions in the second integral about
$\mbox {\boldmath$r$}(0)$, we obtain
\begin{eqnarray}
\label{eq: 11.3} \delta r_\mu(\Delta t)-\delta r_\mu(0)&=&
{1\over{m}}\sum_\nu \int_0^{\Delta t}{\rm d}t_1\
\int_{-\infty}^{t_1}{\rm d}t_2\ \exp[-\gamma(t_1-t_2)]\,F_{\mu \nu
}(\mbox{\boldmath$r$}(0),t_2)\delta r_\nu (0)
\nonumber \\
&&\hspace*{-2cm}+{1\over{m^2}}\sum_{\nu\lambda}
\int_0^{\Delta t}{\rm d}t_1\int_{-\infty}^{t_1} {\rm d}t_2 \
\exp[-\gamma(t_1-t_2)]\int_0^{t_2} {\rm d}t_3 \int_{-\infty}^{t_3}
{\rm d}t_4\ \exp[-\gamma(t_3-t_4)]
\nonumber \\
&&\hspace*{-2cm}\times\biggl[{\partial^2 f_\mu\over{\partial r_\nu\partial
r_\lambda}}
(\mbox{\boldmath$r$}(0),t_2)\,f_\lambda(\mbox{\boldmath$r$}(0),t_4)
+{\partial f_\mu\over{\partial r_\nu}}
\bigl(\mbox{\boldmath$r$}(0),t_2\bigr)\, {\partial
f_\nu\over{\partial r_\lambda}}
(\mbox{\boldmath$r$}\bigl(0),t_4\bigr)\biggr] \delta
r_\nu(0)\delta r_\lambda(0)
\nonumber \\
&&\hspace*{-2cm}+O(f^3)\ .
\end{eqnarray}
This approximation is valid when the absolute displacement $\Delta
\mbox{\boldmath$r$}$ during time $\Delta t$ is small compared to
the correlation length: $\vert \Delta \mbox{\boldmath$r$}(\Delta
t)\vert\ll \xi$. The force is derived from potentials, as
specified by equation (\ref{eq: 3.5}). The potentials $\phi $ and
$A_\mu$ have statistics (given by (\ref{eq: 3.6})) which are
spatially homogeneous and isotropic. Consider the correlation
function of two such quantities $A$ and $B$ which have spatially
homogeneous statistics. The following relation holds between
correlation functions involving derivatives:
\begin{equation}
\label{eq: 11.6}
\biggl\langle {\partial A\over{\partial r_\lambda}}(\mbox{\boldmath$r$},t)
B(\mbox{\boldmath$r$}',t)+A(\mbox{\boldmath$r$},t)
{\partial B\over{\partial r'_\lambda}}(\mbox{\boldmath$r$}',t')
\biggr\rangle=0
\ .
\end{equation}
Applying this relation to (\ref{eq: 11.3}), with $\partial
f_\mu/\partial r_\nu$ and $f_\nu$ playing the role of the
functions $A$ and $B$, we see that
\begin{equation}
\label{eq: 11.7}
\langle \delta \mbox{\boldmath$r$}(\Delta t)-\delta \mbox{\boldmath$r$}(0)
\rangle=O(\Delta t^2)
\ .
\end{equation}

\subsubsection*{Variance of displacement}

The variance of (\ref{eq: 11.1}) is
\begin{equation}
\label{eq: 11.8}
\langle [\delta \mbox{\boldmath$r$}(\Delta t)-\delta \mbox{\boldmath$r$}(0)]^2
\rangle =\Delta t{1\over {m^2}}\int_{-\infty}^\infty {\rm d}t\
\langle \delta \mbox{\boldmath$p$}(t)\cdot \delta \mbox{\boldmath$p$}(0)\rangle
+O(\Delta t^2)\,.
\end{equation}
The correlation function of the components of the momentum difference is
\begin{eqnarray}
\label{eq: 11.9} \langle \delta p_\mu(t)\delta p_{\mu'}(0)\rangle
&\sim &\sum_{\nu\nu'}\int_{-\infty}^0{\rm d}t_1
\int_{-\infty}^0{\rm d}t_2\ \exp[\gamma(t_1+t_2)] \nonumber \\
&\times& \biggl\langle {\partial f_\mu\over{\partial
r_\nu}}(\mbox{\boldmath$r$}(t+t_1),t+t_1) {\partial
f_{\mu'}\over{\partial
r_{\nu'}}}(\mbox{\boldmath$r$}(t_2),t_2)\biggr\rangle \,\delta
r_\nu(0) \delta r_{\nu'}(0) \ .
\end{eqnarray}
We consider the evaluation of the momentum correlation function
(\ref{eq: 11.9}) in successively the underdamped and overdamped
limits. In both cases we assume that the displacement $\Delta
\mbox{\boldmath$r$}$ during the relevant correlation time is small
compared to $\xi$. In the underdamped case, $\Omega \ll 1$, we can
approximate the correlation function appearing in (\ref{eq: 11.9})
by a delta function multiplied by a weight
\begin{equation}
\label{eq: 11.10} \biggl\langle {\partial f_\mu\over{\partial
r_\nu}}(\mbox{\boldmath$r$}(t_1),t_1) {\partial
f_{\mu'}\over{\partial
r_{\nu'}}}(\mbox{\boldmath$r$}(t_2),t_2)\biggr\rangle \sim
2\,{\cal D}_{\mu\nu,\mu'\nu'}\,\delta (t_1-t_2)
\end{equation}
where the coefficients ${\cal D}_{\mu\nu,\mu'\nu'}$ are defined by
equation (\ref{eq: 3.2}).
Using this approximation to simplify (\ref{eq: 11.9})
we find
\begin{equation}
\label{eq: 11.11} \langle \delta p_\mu(t)\delta p_{\mu'}(0)\rangle
={1\over \gamma}\exp(-\gamma \vert t\vert) \sum_{\nu\nu'}{\cal
D}_{\mu\nu,\mu'\nu'}\,\delta r_\nu(0)\delta r_{\nu'}(0) \ ,\ \ \
{\rm (underdamped\ case)},\ \Omega      \ll 1 \ .
\end{equation}
In the case where $\gamma \tau\gg 1$, the integral (\ref{eq: 11.9})
is dominated by the region $t_1\sim t_2\sim 0$, and we obtain
\begin{equation}
\label{eq: 11.12} \langle \delta p_\mu(t) \delta
p_{\mu'}(0)\rangle =\sum_{\nu\nu'}{1\over{\gamma^2}} \biggl\langle
{\partial f_\mu\over{\partial r_\nu}}(\mbox{\boldmath$r$}(t),t)
{\partial f_{\mu'}\over{\partial
r_{\nu'}}}(\mbox{\boldmath$r$}(0),0)\biggr\rangle \,\delta r_\nu
(0)\delta r_{\nu'}(0) \ ,\ \ \ {\rm (overdamped\ case)},\ \Omega
\gg 1 \ .
\end{equation}
Thus we find that the momentum correlation functions are different
in the underdamped and overdamped cases. However, when we evaluate
the variance of the change in spatial separation using (\ref{eq: 11.8}),
we find the same result in both limits:
\begin{equation}
\label{eq: 11.13}
\langle [\delta \mbox{\boldmath$r$}(\Delta t)
-\delta \mbox{\boldmath$r$}(0)]^2\rangle
={2\Delta t\over{m^2\gamma^2}}\sum_{\mu\nu\nu'}
{\cal D}_{\mu\nu,\mu\nu'}\,\delta r_\nu(0)\delta r_{\nu'}(0)
\ .
\end{equation}
Equations (\ref{eq: 11.7}) and (\ref{eq: 11.13}) give the
statistics of the change in the particle separation after a short
time $\Delta t$. This result is valid both when $\Omega \gg 1$
(overdamped limit) and when $\Omega \ll 1$ (underdamped limit),
provided that the particle displacement is sufficiently small . In
both cases we estimate the displacement as $\vert\Delta
\mbox{\boldmath$r$}(\Delta t)\vert\approx \sqrt{D\Delta t}$, where
$D\approx u^2\tau$ is the spatial diffusion constant. In the
underdamped case the correlation time is $O(1/\gamma)$, and the
requirement that $\Delta \mbox{\boldmath$r$}(\gamma^{-1})\ll \xi$
gives the condition $u\sqrt{\tau}/\xi\sqrt{\gamma}\ll 1$ that is
$\epsilon \ll 1$. In the overdamped limit (\ref{eq: 11.7}) and
(\ref{eq: 11.13}) are always valid when $\kappa\ll 1$.

\subsubsection*{Relation to advective model}

Equations (\ref{eq: 11.7}) and (\ref{eq: 11.13}) imply that both
in the overdamped limit and in the underdamped limit when
$\epsilon\ll 1$, the dynamics may be approximated by a random
advection model. In this model the displacement of a particle
during time $\Delta t$ is a function only of the position of the
particle at time $t$, and not its momentum: we write the
displacement at $t=n\Delta t$ as
\begin{equation}
\label{eq: 7.1}
\mbox{\boldmath$r$}(t+\Delta t)=\mbox{\boldmath$r$}(t)+
\mbox{\boldmath$w$}_n(\mbox{\boldmath$r$})\Delta t
\end{equation}
The values of $\bbox{w}_n$ at successive time steps are uncorrelated
and average to zero, but
$\bbox{w}_n(\bbox{r})$ is a smoothly varying function of position, so that
nearby particles follow similar paths in this random advection
model. The spatial correlation of $\bbox{w}_n(\bbox{r})$ at equal
times ($n=n'$) is determined by that of $\bbox{u}(\bbox{r},t)$.

The separation $\delta \mbox{\boldmath$r$}$ of two
trajectories satisfies
\begin{equation}
\label{eq: 7.2}
\delta \mbox{\boldmath$r$}(t+\Delta t)=[{\bf I}+{\bf M}_n]
\delta\mbox{\boldmath$r$}(t)
\end{equation}
where ${\bf I}$ is the identity matrix and where ${\bf M}_n$ is a
matrix with elements $({\bf M}_n)_{\mu\nu}=(\partial
(\mbox{\boldmath$w$}_n)_\mu/\partial r_\nu)\delta t$. The elements
of this matrix are assumed to have mean value zero and to have
diffusive increments: for consistency with (\ref{eq: 11.13}) we
write
\begin{equation}
\label{eq: 7.3} \langle ({\bf M}_n)_{\mu\nu}\rangle=0
\end{equation}
\begin{equation}
\label{eq: 7.4} \langle ({\bf M}_n)_{\mu\nu}({\bf
M}_{n'})_{\mu'\nu'}\rangle={2\over{m^2\gamma^2}} \delta_{nn'}{\cal
D}_{\mu\nu,\mu'\nu'}\Delta t \ .
\end{equation}
Equations (\ref{eq: 7.1}) to (\ref{eq: 7.4}) define an It\^ o type
stochastic process. It is a surprising feature that an It\^ o
rather than a Stratonovich type equation \cite{vKa92} arises as a
stochastic approximation to a system of continuous differential
equations.

Note that we do not claim that the particles are always simply
advected by the flow $\mbox{\boldmath$u$}(\mbox{\boldmath$r$},t)$
when $\Omega\ll 1$ and $\epsilon \ll 1$: in the overdamped case
this is true, but in the underdamped case they are not advected.
The statement is that the particle separations behave as if the
particles were being advected, according to the simple model
(\ref{eq: 7.1}).

\subsection{Lyapunov exponents of Brownian advection model}
\label{sec: 7.2}

The Lyapunov exponents for processes of the type defined by
(\ref{eq: 7.1}) to (\ref{eq: 7.4}) were obtained by LeJan
\cite{LeJ85}. Here we calculate these Lyapunov exponents using a
different and much simpler approach, facilitating direct
comparison with the coefficients in equation (\ref{eq: 6.7}). 
\subsection*{First Lyapunov exponent}

The Lyapunov exponent $\lambda_1$ is obtained by considering the
magnitude of the separations $\delta r(t)=\vert \delta
\mbox{\boldmath$r$}(t)\vert$, at successive time steps:
\begin{equation}
\label{eq: 7.5}
\lambda_1={1\over{\Delta t}}
\biggl\langle\log_{\rm e}
\biggl({\delta r(t+\Delta t)\over{\delta r(t)}}\biggr)\biggr\rangle
\ .
\end{equation}
From (\ref{eq: 7.2}) we obtain (from here on we drop the
subscript labeling the time step on the matrix ${\bf M}_n$)
\begin{equation}
\label{eq: 7.6}
\delta r^2(t+\Delta t)=\delta r^2(t)\,{\bf n}_1\cdot
({\bf I}+{\bf M}+{\bf M}^T+{\bf M}^T{\bf M})\,{\bf n}_1
\end{equation}
where ${\bf n}_1$ is a unit vector such that
$\delta \mbox{\boldmath$r$}=\delta r{\bf n}_1$.
Anticipating that two other unit vectors will be defined
in due course, and writing
\begin{equation}
\label{eq: 7.7}
M_{\mu\nu}'={\bf n}_\mu\cdot {\bf M}\,{\bf n}_\nu
\end{equation}
we obtain
\begin{eqnarray}
\label{eq: 7.8}
{\delta r(t+\Delta t)\over{\delta r(t)}}&=&
\sqrt{1+2M'_{11}+({\bf M}^T{\bf M})'_{11}}
\nonumber \\
&=&1+M'_{11}-{\textstyle{1\over 2}}{M'}_{11}^2 +{\textstyle{1\over
2}}({\bf M}^T{\bf M})'_{11}+O(M^3)
\nonumber \\
\log_{\rm e}\biggl({\delta r(t+\Delta t)\over{\delta r(t)}}\biggr)&=&
M'_{11}-{M'}_{11}^2+{\textstyle{1\over 2}}({\bf M}^T{\bf M})'_{11}+O(M^3)
\nonumber \\
&=&M'_{11}+{\textstyle{1\over 2}}\sum_\mu {M'}_{1\mu }^2-{M'}_{11}^2+O(M^3)
\ .
\end{eqnarray}
Taking the expectation value using (\ref{eq: 7.3}) and
(\ref{eq: 7.4}) we obtain
\begin{equation}
\label{eq: 7.9} \lambda_1={1\over {m^2\gamma^2}}\biggl[
\sum_{\mu=1}^d {\cal D}_{\mu 1 \mu 1}-2{\cal D}_{1111}\biggr]
\end{equation}
where $d$ is the number of space dimensions.
Finally we relate this expression for the Lyapunov exponent
to the results of section \ref{sec: 6}, showing agreement
with the lowest-order terms of the expansions $\epsilon$.
Using the properties of the elements ${\cal D}_{\mu \nu\mu'\nu'}$
discussed in section \ref{sec: 3}, (\ref{eq: 7.9}) gives
\begin{equation}
\label{eq: 7.20} \lambda_1=\gamma {{\cal
D}_1\over{m^2\gamma^3}}(2\Gamma -1) =\gamma \epsilon^2 (2\Gamma-1)
\end{equation}
(here we used the fact that ${\cal D}_{1313}={\cal D}_{1212}\equiv
{\cal D}_2 \equiv \Gamma {\cal D}_1$). This agrees with the
leading-order term in the expansion (\ref{eq: 6.6}), (\ref{eq:
6.7}).

\subsection*{Second Lyapunov exponent}

For the second Lyapunov exponent we
follow an approach analogous to that of section \ref{sec: 2}:
we consider two vectors $\delta \mbox{\boldmath$r$}_1=\delta r{\bf n}_1$
and $\delta \mbox{\boldmath$r$}_2=\delta \mbox{\boldmath$r$}_1+
\delta r \delta\theta {\bf n}_2$, where ${\bf n}_2$ is orthogonal
to ${\bf n}_1$ and where $\delta\theta \ll 1$. A third unit vector is defined by
${\bf n}_3={\bf n}_1\wedge {\bf n}_2$. The area of the
parallelogram spanned by the vectors $\delta \mbox{\boldmath$r$}_1$
and $\delta \mbox{\boldmath$r$}_2$ is
$\delta {\cal A}(t)=
\vert \delta\mbox{\boldmath$r$}_1(t)\wedge
\delta\mbox{\boldmath$r$}_2(t)\vert $.
The second Lyapunov exponent may be obtained from the relation
\begin{equation}
\label{eq: 7.10}
\lambda_1+\lambda_2={1\over{\Delta t}}\biggl\langle\log_{\rm e}
\biggl({\delta{\cal A}(t+\Delta t)\over
{\delta{\cal A}(t)}}\biggr)\biggr\rangle
\ .
\end{equation}
We find $\delta{\cal A}(t)=\delta r^2\delta\theta$ and
\begin{equation}
\label{eq: 7.11}
\delta{\cal A}(t+\Delta t)=
\vert {\bf n}_1\wedge{\bf n}_2+\mbox {\boldmath$z$}\vert
\end{equation}
where
\begin{equation}
\label{eq: 7.12}
\mbox{\boldmath$z$}={\bf M}{\bf n}_1\wedge {\bf n}_2
+{\bf n}_1\wedge {\bf M}{\bf n}_2+{\bf M}{\bf n}_1\wedge {\bf M}{\bf n}_2
\ .
\end{equation}
This implies
\begin{equation}
\label{eq: 7.13}
\left(\frac{\delta{\cal A}(t+\Delta t)}{\delta{\cal A}(t)}\right)^2=1+2{\bf n}_3
\cdot \mbox{\boldmath$z$}
+\mbox{\boldmath$z$}\cdot \mbox{\boldmath$z$}
\ .
\end{equation}
It is convenient to express these vectors in terms of components:
for example we have
${\bf M}{\bf n}_1=M'_{11}{\bf n}_1+M'_{21}{\bf n}_2+M'_{31}{\bf n}_3$
so that ${\bf M}{\bf n}_1\wedge {\bf n}_2=M'_{11}{\bf n}_3-M'_{31}{\bf n}_1$.
We find
${\bf n}_3\cdot \mbox{\boldmath$z$}=M'_{11}+M'_{22}+M'_{11}M'_{22}
-M'_{21}M'_{12}$ and
$\mbox{\boldmath$z$}\cdot \mbox{\boldmath$z$}={M'}_{31}^2+{M'}_{32}^2
+{M'}_{11}^2+{M'}_{22}^2+2M'_{11}M'_{22}+O(M^3)$. This gives
\begin{eqnarray}
\label{eq: 7.14}
\left(\frac{\delta{\cal A}(t+\Delta t)}{\delta{\cal A}(t)}\right)^2
&=&1+2(M'_{11}+M'_{22})+2M'_{11}M'_{22}
-2M'_{21}M'_{12}
\nonumber \\
&&+{M'}_{31}^2+{M'}_{32}^2+(M'_{11}+M'_{22})^2+O(M^3)
\nonumber \\
{\delta{\cal A}(t+\Delta t)\over{\delta{\cal A}(t)}}&=&
1+M'_{11}+M'_{22}-M'_{21}M'_{12}+  M'_{11} M'_{22}+
{\textstyle{1\over 2}}{M'}_{31}^2+{\textstyle{1\over 2}}{M'}_{32}^2
+O(M^3)
\nonumber \\
\log_{\rm e}{\delta{\cal A}(t+\Delta t)\over {\delta{\cal A}(t)}}&=&
M'_{11}+M'_{22}-M'_{12}M'_{21}
-{\textstyle{1\over 2}}{M'}_{11}^2-{\textstyle{1\over 2}}{M'}_{22}^2
+{\textstyle{1\over 2}}{M'}_{31}^2+{\textstyle{1\over 2}}{M'}_{32}^2\\
&&+O(M^3)\nonumber
\ .
\end{eqnarray}
Using (\ref{eq: 7.10}), (\ref{eq: 7.3}) and (\ref{eq: 7.4}), the sum
of the first two Lyapunov exponents is found to be
\begin{equation}
\label{eq: 7.15}
\lambda_1+\lambda_2={1\over{m^2\gamma^2}}\bigl({\cal D}_{3131}+{\cal D}_{3232}
-{\cal D}_{1111}-{\cal D}_{2222}-2{\cal D}_{2112}\bigr)
\ .
\end{equation}
Expressing (\ref{eq: 7.15}) in terms of the notation used in
section \ref{sec: 6}, we find
\begin{equation}
\label{eq: 7.21} \lambda_1+\lambda_2
={1\over{m^2\gamma^2}}\bigl(2{\cal D}_2-2{\cal D}_1-2{\cal
D}_3\bigr) =\gamma {{\cal D}_1\over{m^2\gamma^3}}(2\Gamma -2
-2\sigma) =\gamma \epsilon^2(3\Gamma -3) \ .
\end{equation}
This gives $\lambda_2=\gamma \epsilon^2(\Gamma-2)$, which is
consistent with the leading-order
term of the expansion of $\lambda_2$ in (\ref{eq: 6.7}).

\subsection*{Third Lyapunov exponent}

The third Lyapunov exponent is determined by considering the sum
of the first three Lyapunov exponents, which are obtained from the
mean logarithmic derivative of the volume element $\delta {\cal
V}(t)$:
\begin{equation}
\label{eq: 7.16}
\lambda_1+\lambda_2+\lambda_3={1\over{\Delta t}}
\biggl\langle \log_{\rm e}\biggl({{\cal V}(t+\Delta t)
\over{\cal V}(t)}\biggr)\biggr\rangle
\ .
\end{equation}
We have
\begin{eqnarray}
\label{eq: 7.17}
{{\cal V}(t+\Delta t)\over{\cal V}(t)}&=&{\rm det}({\bf I}+{\bf M})
\nonumber \\
\log_{\rm e}\biggl({{\cal V}(t+\Delta t)\over{\cal V}(t)}\biggr)&=
&\log_{\rm e}{\rm det}({\bf I}+{\bf M}) ={\rm tr}\log_{\rm e}({\bf
I}+{\bf M})
\nonumber \\
&=&{\rm tr}\bigl[{\bf M}-{\textstyle{1\over 2}}{\bf M}^2\bigr]+O(M^3)
\ .
\end{eqnarray}
The sum of the first three Lyapunov exponents is therefore
\begin{eqnarray}
\label{eq: 7.18}
\lambda_1+\lambda_2+\lambda_3&=&-{\textstyle{1\over 2}}{1\over {\Delta t}}
\biggl\langle {\rm tr}({\bf M}^2)\biggr\rangle
\nonumber \\
&=&-\frac{3}{m^2\gamma^2}\bigl[{\cal D}_{1111}+{\cal D}_{1221}
+{\cal D}_{1331}\bigr]\,.
\end{eqnarray}
In the notation of section \ref{sec: 6}, (\ref{eq: 7.18}) gives
\begin{equation}
\label{eq: 7.22} \lambda_1+\lambda_2+\lambda_3
={-3\over{m^2\gamma^2}}\bigl({\cal D}_1+2{\cal D}_3\bigr)
=-3\gamma \epsilon^2(1+2\sigma)=-3(2-\Gamma)\gamma \epsilon^2
\end{equation}
Subtracting (\ref{eq: 7.21}) we find $\lambda_3=-3\gamma \epsilon^2$,
which agrees with the leading term of (\ref{eq: 6.7}).

Equation (\ref{eq: 7.18}) may also be written in the form
\begin{eqnarray}
\label{eq: 7.19}
\lambda_1+\lambda_2+\lambda_3&=&-\frac{1}{2}
\int_{-\infty}^\infty {\rm d}t\, \langle
{\rm div}\,\bbox{w}(\bbox{0},t)\,
{\rm div}\,\bbox{w}(\bbox{0},0)\rangle
\ .
\end{eqnarray}
Here $\bbox{w}(\bbox{r},t)$ is related to the velocity field
$\bbox{w}_n(\bbox{r})$ in equation (\ref{eq: 7.1}) by
$\bbox{w}_n(\bbox{r}) = \bbox{w}(\bbox{r},n \Delta t)$. A similar
expression has been quoted by Balkovsky {\sl et al.} \cite{Bal01}.

\section{Correction to overdamped limit}
\label{sec: 8}

In the previous section, the advective approximation of the
solution of equation (\ref{eq: 1.1}) was considered. We showed
that whenever $\epsilon \ll 1$, at leading order in $\kappa $ we
can apply a model in which the particles are advected by a random
flow. The only case where the advective approximation is
inadequate in the overdamped limit is the following. Consider the
quantity $\Delta$, defined in equation (\ref{eq: 0.0}). Its sign
and magnitude determine whether or not clustering occurs, and to
which extent. In the case of an incompressible flow, $\Delta$
vanishes in the advective approximation, because the flow is
volume preserving (this also follows from equation (\ref{eq:
7.22}), noting that $\Gamma=2$ for incompressible flow). In order
to compute whether $\Delta$ is positive or negative (which
determines whether or not clustering happens) it is thus necessary
to go beyond the advective approximation, as explained in the
following.

We consider an incompressible flow field $\bbox{u}(\bbox{r},t)$.
We make use of a result due to Maxey \cite{Max87} who has derived
a \lq synthetic' velocity field including a correction term which
is proportional to $1/\gamma$ and shown that this modified field
describes, to leading order, corrections to the purely advective
limit. Maxey's synthetic field was subsequently employed in
\cite{Elp96,Pin99,Bal01} to study particle-density fluctuations in
incompressible fluids in the near-advective limit.

As pointed out by Maxey, this correction term gives rise to a
potential component in the synthetic advection field which,
according to equation (\ref{eq: 7.18}), implies that $\Delta>0$.
The aim of this section is to evaluate this correction, and to
express it in terms of the dimensionless parameters $\kappa$ and
$\Omega$.

Integrating (\ref{eq: 1.1}) one obtains
\begin{eqnarray}
\label{eq: 8.1}
\bbox{p}(t)  &=& \int_{-\infty}^t\!\!{\rm d}t_1\ \exp[-\gamma(t-t_1)]
\bbox{f}(\bbox{r}(t_1),t_1)
\nonumber \\
&=& \frac{1}{\gamma}\Big[\bbox{f}(\bbox{r}(t),t)-
\int_{-\infty}^t\!\!{\rm d}t_1\ \exp[-\gamma(t-t_1)] \frac{\rm
d}{{\rm d}t_1}\bbox{f}(\bbox{r}(t_1),t_1)\Big]
\end{eqnarray}
ignoring an initial transient. Using $\bbox{f} = m\gamma \bbox{u}$
this gives, to leading order in $\Omega^{-1}=(\gamma\tau)^{-1}$
\begin{equation}
\label{eq: 8.3} \dot{\bbox{r}} =
\bbox{u}(\bbox{r}(t),t)-\frac{1}{\gamma}
\Big[\frac{\partial}{\partial t}\bbox{u}(\bbox{r}(t),t)
+\big(\bbox{u}(\bbox{r}(t),t)\cdot
\bbox{\nabla}\big)\dot{\bbox{r}}\Big]\ .
\end{equation}
As pointed out by Maxey (equation (5.9) in \cite{Max87}) this
equation describes advection in a synthetic velocity field
\begin{equation}
\label{eq:syn} \bbox{v} = \bbox{u}-\frac{1}{\gamma}
\Bigg[\frac{\partial \bbox{u}}{\partial t} +\big(\bbox{u}\cdot
\bbox{\nabla}\big)\bbox{u}\Bigg] \ .
\end{equation}
The sum of the Lyapunov exponents for this advection problem are
given by (\ref{eq: 7.19}). Since $\mbox{div}\,\bbox{v}
=\gamma^{-1} \bbox{\nabla}(\bbox{u}\cdot \bbox{\nabla})\bbox{u}$,
we have
\begin{equation}
\lambda_1+\lambda_2+\lambda_3=-\frac{1}{2\gamma^2}
\int_{-\infty}^\infty {\rm d}t\, \Big\langle
\Big(\bbox{\nabla}\cdot(\bbox{u}\cdot\bbox{\nabla})\bbox{u}\Big)_t
\Big(\bbox{\nabla}\cdot(\bbox{u}\cdot\bbox{\nabla})\bbox{u}\Big)_0
\Big\rangle \ .
\end{equation}
We will estimate this expression for a particular example, where
the correlation function $C(R,t)$ in (\ref{eq: 3.6}) factorises
into a spatial and a time-dependent part: $C(R,t) = C_0(R)
\Phi(t)$. In the incompressible case where $\nabla\cdot
\mbox{\boldmath$u$}=0$, we obtain
\begin{equation}
\lambda_1+\lambda_2+\lambda_2 = -\left\langle\left(\sum_{\mu\nu}
\frac{\partial u_\mu}{\partial r_\nu} \frac{\partial
u_\nu}{\partial r_\mu}\right)^2\right\rangle\,
\frac{1}{2\gamma^2}\int_{-\infty}^\infty {\rm d}t\,\Phi^2(t)\ .
\end{equation}
In terms of the characteristic quantities of the flow, we
therefore estimate $\lambda_1+\lambda_2+\lambda_3 \sim
\kappa^4/(\Omega^2\tau)$. Together with  (\ref{eq: 7.20}) this
gives, for $\Omega \gg 1$,
\begin{equation}
\label{eq: d1} \Delta \sim\frac{\kappa^2}{\Omega^2}\ .
\end{equation}
This result should be contrasted with (\ref{eq:sum}), valid for
$\Omega \ll 1$, which implies that $\Delta \sim \kappa^2/\Omega$.
Thus $\Delta$ is given in terms if different combinations of
dimensionless parameters $\kappa$ and $\Omega$ for $\Omega\ll 1$
and $\Omega \gg 1$. In the underdamped case, $\Delta$ is obtained
as a perturbation expansion in $\epsilon = \kappa \Omega^{-1/2}$.
In the overdamped case, by contrast, it is obtained as a
perturbation expansion in $\kappa/\Omega$ and $\kappa/\Omega^2$.
We note that expression (\ref{eq: d1}) is always small in the
domain of its validity: $\Omega \gg 1$ and $\kappa <  1$. The
results of section \ref{sec: 6} by contrast, remain valid when
$\Delta $ is of order unity.

\section{Concluding remarks}
\label{sec:13}

\subsection{Conclusions}
\label{sec: 13.1}

In this paper we have characterised the local properties of the
motion of inertial particles in a short-time correlated random
flow model. We calculated the Lyapunov exponents using an exact
series expansion, and characterised the rate of caustic formation
in terms of an escape process.

The results give important   new insights into the mechanism for
particle clustering. The relevant dimensionless parameter is
$\epsilon \sim \kappa \Omega^{-1/2}$, and we find that the
dimension deficit is significant when $\epsilon $ is of order
unity, reaching a maximum value of $\Delta_{\rm max}\approx 0.35$
at $\epsilon_{\rm max}\approx 0.21$. This contradicts the accepted
explanation for particle clustering, based on the centrifuge
mechanism, because (when $\kappa\ll 1$) it implies that clustering
onto a fractal set can occur when the Stokes number $\Omega^{-1}$
is large. For fully developed turbulent flow, we have
$\kappa=O(1)$ so that $\epsilon\sim \Omega^{-1/2}$, and in that
case our theory does predict that clustering onto a fractal set
occurs when the Stokes number is of order unity. This is
consistent with numerical experiments. We cannot exclude the
possibility that the centrifuge effect has some relevance, however
we note  that our theory predicts  $\Delta_{\rm max} = 0.35$ for
the maximal dimension deficit, see figure \ref{fig:f1}{\bf b},
also \cite{Dun05}. The value obtained from direct numerical
simulations of a Navier-Stokes flow is very similar, $\Delta_{\rm
max} = 0.4$ \cite{Bec06}.

For $\epsilon>\epsilon_{\rm c}\approx 0.33$, the dimension deficit
is zero, and the particles will not cluster onto a fractal set.
However, as $\epsilon$ increases the rate of caustic formation
increases. The particle density can diverge on caustic lines, as
described in \cite{Wil05}.

\begin{figure}
\centerline{\includegraphics[width=8cm]{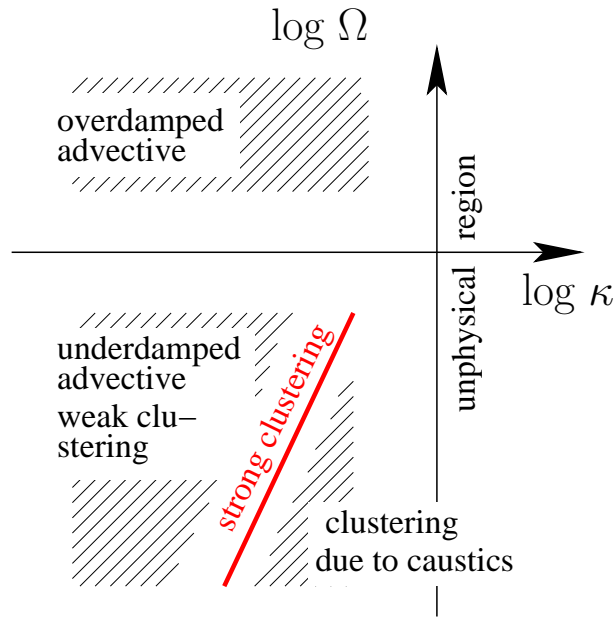}}
\caption{\label{fig:f7} Schematic diagram showing how the
behaviour of particles suspended in an incompressible fluid
depends upon the parameters $\kappa $ and $\Omega $. }
\end{figure}

Figure 7 is a schematic illustration of how the unmixing  of
particles suspended in an incompressible fluid (that is a flow
with $\Gamma=2$) depends upon the parameters $\kappa$ and
$\Omega$. Instead of clustering being confined to $\Omega \sim 1$,
it may occur on a ray in logarithmic parameter space extending
into the underdamped region. The solid line in figure
\ref{fig:f7}, $\epsilon \sim \kappa \Omega^{-1/2} =
\mbox{const}.$, indicates schematically the critical line where
$\Delta$ reaches zero. Above the solid line $\Delta$ is always
positive, but tends to zero for small $\epsilon$ as $\Delta =
10\epsilon^2 \sim \kappa^2/\Omega$. In section 8 we showed that in
the limit of $\epsilon\rightarrow 0$ the dynamics behaves like
advection, even if $\Omega\ll 1$.

\subsection{Experimental considerations}
\label{sec: 13.2}

We make some final comments concerning the situations in which
clustering should be observable. The velocity field of
fully-developed turbulent flow has a power-law energy spectrum,
with an upper and lower cutoff \cite{Fri97}. The smaller
length scale is the Kolmogorov length, defined by
$\eta=(\nu^3/{\cal E})^{1/4}$ where $\nu$ is the kinematic
viscosity and ${\cal E}$ is the rate of dissipation per unit mass
of fluid. This is the size of the smallest vortices generated by
the turbulence. In section \ref{sec: 3.3} we show that it is this
small length scale $\eta$ which corresponds to the correlation
length $\xi$ in our theory. Our results indicate that clustering
can certainly occur on a scale smaller than $\xi$, but what
happens on larger length scales is less certain.

\subsubsection*{Lengthscale for clustering}

It is not entirely clear whether particles in turbulent flows can
exhibit clustering effects on length scales greater than the
Kolmogorov length, $\eta$. Particles separated by distances
greater than $\eta$ will be separated by Richardson diffusion
\cite{Ric26}, but that is not necessarily incompatible with some
type of clustering. Numerical evidence for such an effect has been
presented \cite{Gam04}, but the range of Stokes and Reynolds
numbers are not sufficiently large to allow clear conclusions. At
very small Stokes number, suspended particles are advected and
must be mixed to uniform concentration in a turbulent flow. At
large Stokes number, however, it is conceivable that particles
could be centrifuged away from vortices with length scale $L\gg
\eta$ when their rotation period $t_L$ is such that the
length-dependent Stokes number ${\rm St}_L=1/(\gamma T_L)$ is of
order unity \cite{Fal03}. The Kolmogorov theory gives $t_L\sim
(L^2/{\cal E})^{1/3}$. Consequently this picture implies
clustering on a length scale $L\sim ({\cal E}/\gamma^3)^{1/2}$.

However, there is an argument which suggests that such large-scale
clusters might not form. At large Stokes number, the formation of
numerous caustics implies that the velocity field of the suspended
particles is multi-valued, so that in the limit ${\rm St}\to
\infty$ the suspended particles are expected to behave as a gas
\cite{Abr75}. Particles at the same position move with relative
speed $\Delta v$; using the Kolmogorov cascade principle, we
estimate that the variance of $\Delta v$ is $\langle \Delta
v^2\rangle \sim {\cal E}/\gamma$ \cite{Meh06}. The motion of the
suspended particles is damped at a rate $\gamma$, so that they
travel in an approximately straight line for a distance $\Delta v
/\gamma$. We see that $\Delta v/\gamma \sim ({\cal
E}/\gamma^2)^{1/2}\sim L$. So particles within a cluster of size
$L$ travel distances of order $L$ in random directions before
their relative motion is damped out. This argument suggests that
clustering may be difficult to observe on scales larger than the
Kolmogorov length.

\subsubsection*{Effects of gravity}

The Kolmogorov scale is rather insensitive to the value of ${\cal
E}$: for $\nu \sim 10^{-5}{\rm m}^2{\rm s}^{-1}$ (air at standard
atmospheric conditions) and ${\cal E} \sim 1 {\rm m}^2{\rm
s}^{-3}$, we have $\eta \sim 10^{-4}{\rm m}$. Our results
concerning the Lyapunov exponents are relevant to clustering of
particles on length scales below the Kolmogorov length and
therefore describe clustering on very small length scales. They
would be relevant to the aggregation of aerosol particles, but it
seems unlikely that it could explain the macroscopic clustering
observed in the experiments of Fessler and Eaton and \cite{Fes94}.

In discussing the conditions under which the clustering effect is
observable, we must also consider the effects of gravity.
Gravitational effects may be assumed negligible when motion at the
terminal velocity moves particles by a distance which is small
compared to $\xi$ in time $\tau$. Noting that the terminal
velocity is $g/\gamma$, this condition may be written
$g\tau^2/\xi\Omega\ll 1$. For fully developed turbulence, using
the Kolmogorov theory we estimate
\begin{equation}
\label{eq: 0.3} \frac{g\tau^2}{\xi}\sim\frac{g\nu^{1/4}}{{\cal
E}^{3/4}}\ .
\end{equation}
This quantity is small only for highly energetic turbulence. Thus,
in terrestrial applications, direct application of the model to
three-dimensional turbulent flows requires very intense turbulence
if the effects of gravity are to be neglected.

{\em Acknowledgments}. BM acknowledges financial support
from Vetenskapsr\aa{}det and from the research platform
\lq Nanoparticles in an interactive environment' at G\"oteborg
university.

\newpage


\end{document}